\documentclass[%
 reprint,
 superscriptaddress,
nofootinbib,
 amsmath,amssymb,
 aps,
pra,
onecolumn,
notitlepage,
]{revtex4-1}

\usepackage{graphicx}
\usepackage{dcolumn}
\usepackage{subcaption}
\usepackage{bm}
\usepackage{breakurl}
\usepackage{enumitem}

\begin{document}

\title{Criticality of Spin Systems with Weak Long-Range Interactions}

\author{Nicol\`o Defenu}
\address{Institut f\"ur Theoretische Physik, Universit\"at 
Heidelberg, D-69120 Heidelberg, Germany}

\author{Alessandro Codello}
\address{Department of Physics, Southern University of Science and Technology,
  Shenzhen 518055, China}

\author{Stefano Ruffo}
\address{SISSA, Via Bonomea 265, I-34136 Trieste, Italy}
\address{INFN, Sezione di Trieste, I-34151 Trieste, Italy}

\author{Andrea Trombettoni}
\address{CNR-IOM DEMOCRITOS Simulation Center, Via Bonomea 265, 
I-34136 Trieste, Italy}
\address{SISSA, Via Bonomea 265, I-34136 Trieste, Italy}
\address{INFN, Sezione di Trieste, I-34151 Trieste, Italy}


\begin{abstract}
The study of critical properties of systems with long-range interactions
has attracted in the last decades a continuing interest and motivated the
development of several analytical and numerical techniques, in particular in
connection with spin models. From the point of view of the
investigation of their criticality, a special role is played by systems in
which the interactions are long-range enough that their universality class is
different from the short-range case and, nevertheless, they maintain the extensivity
of thermodynamical quantities. Such interactions are often called weak long-range.
In this paper we focus on the study of the critical behaviour of spin systems with weak-long range couplings
using renormalization group, and we review their remarkable properties. For the sake of clarity and
self-consistency, we start from the classical $O(N)$ spin models and we then move to
quantum spin systems.
\end{abstract}
\maketitle
\section{Introduction}
\label{sec:intro}
In this Special Issue several aspects of the equilibrium and dynamical properties of systems
with long-range (LR) interactions are discussed. In the paper we apply concepts and tools developed for LR systems,
including the modern renormalization group approach, to the study 
of spin models with weak LR interactions. Such interactions are able to
modify the universal properties of the systems in which they act, but
anyway preserve the extensivity of the thermodynamic quantities. Our main goal
is to present and apply the formalism of the functional renormalization group
(FRG) to weak LR systems and show the capability 
of the FRG to clarify their critical properties,
which are in many cases still unknown and sometimes diffucult
to obtain with other approaches.

Given the wide framework already presented in the other contributions to this Special
Issue, it is not necessary to extensively emphasize that
LR interactions play an important and paradigmatic
role in the study of many body interacting systems, such as $O(N)$ symmetric models.
Indeed, the importance of LR interactions is motivated by their presence
in several systems ranging from plasma physics to astrophysics 
and cosmology\,\cite{Campa2009,Campa2014}.

In order to understand the typical phenomena occurring in spin
models with power-law
LR couplings and define the weak LR regime, let us first consider the classical
$O(N)$ symmetric models, whose Hamiltonian reads
\begin{equation}
\label{NvectorSystemHamiltonian}
H=-\frac{J}{2}\sum_{i \neq j} 
\frac{\mathbf{S}_{i}\cdot\mathbf{S}_{j}}{|i-j|^{d+\sigma}}\,.
\end{equation}
The spin variables $\mathbf{S}_{i}$ are unit vectors with $N$ components,
placed at the sites, labeled by the index $i$, of a $d$ dimensional lattice.
The coupling constant $J$ is constant for a decay exponent $\sigma>0$,
conversely for $\sigma \leq 0$ the coupling constant $J$ needs to be rescaled
by an appropriate power of the system size to absorb the divergence of the interaction energy density the thermodynamic
limit\,\cite{Campa2009,Campa2014}.

When $\sigma>0$ the model may have a second order phase 
transition. The main result is that three different regimes can
occur as a function of the parameter $\sigma$\,\cite{Fisher1972,Sak1973}:
\begin{itemize}
\item for $\sigma\leq d/2$ the mean--field approximation
  correctly describes the universal behavior; 
\item for $\sigma$ greater than a threshold value, $\sigma_*$, 
the model has the same critical exponents of the short-range (SR) model
(the SR model is obtained in the limit $\sigma \to \infty$); 
\item for $d/2<\sigma \le \sigma_*$ the system exhibits 
peculiar LR critical exponents.
\end{itemize}
Therefore, the weak LR regime is the one with $0<\sigma \le \sigma_{*}$,
where the thermodynamic quantities are extensive and the critical properties
are not the ones of the SR model
[notice that, as we will
  discuss in the following, the SR critical exponents may
  be found
  also at $\sigma=\sigma_\ast$]. Inside the weak LR region, the more
interesting weak LR regime is the one with $d/2<\sigma \leq \sigma_{*}$
where the critical properties are neither SR nor mean-field. 
The $1$-dimensional Ising model ($d=N=1$) deserves a special mention
since for $\sigma=\sigma_{*}=1$\,\cite{Dyson1969,Thouless1969,Anderson1970}
a topological phase transition in the Berezinskii-Kosterlitz-Thouless (BKT) universality
occurs\,\cite{Cardy1981,Frohlich1982,Luijten2001}
(more references can be found in\,\cite{Luijten1997}), with effects of
disorder being studied in\,\cite{Balog2014}. 

For any value of $d$ and $N$, the expression 
$\eta=2-\sigma$ for the critical exponent $\eta$, which characterizes the decay of
the two point correlation function at the critical point, was found in Ref.\,\cite{Fisher1972}
by an $\epsilon$-expansion (at order $\epsilon^2$) and 
conjectured to be exact. This implies a
discontinuity of the anomalous dimension $\eta$ as a function of
the parameter
$\sigma$, when $\sigma$ reaches $\sigma_*$, since $\sigma_*=2$
according to the calculations of Ref.\,\cite{Fisher1972}.
A way out from this scenario was proposed by J. Sak in Ref.\,\cite{Sak1973}, who confirmed the expression $\eta=2-\sigma$
for all $\sigma<\sigma_*$, but found a different threshold value $\sigma_*=2-\eta_{SR}$, where 
$\eta_{SR}$ is the $\eta$ exponent of the SR model. Then, $\eta$ is 
a continuous function of $\sigma$ and
there is no correction to the canonical dimension of the field in the case
of LR interactions. It is is fair to say that
most of the Monte Carlo (MC) results, based on MC 
algorithms specific for LR interactions\,\cite{LUIJTEN1995,Fukui2009,Gori2016}, 
confirmed this picture\,\cite{Luijten2002,Angelini2014,Gori2016,Horita2016}.
Nevertheless, different behaviours compatible with $\sigma_*=2$
were discussed in the literature\,\cite{Suzuki1973,Yamazaki1977,Picco2012,Blanchard2013,Grassberger2013}.
In particular the MC results for a percolation model 
with LR probabilities\,\cite{Grassberger2013} appeared to be
in agreement with the findings of Ref.\,\cite{Blanchard2013} 
and not with Ref.\,\cite{Sak1973}.
At variance, recent results with conformal
bootstrap supported
and clarified the Sak's scenario\,\cite{Paulos2016,Behan2017,Connor2019}.

It is convenient to define the anomalous dimension 
$\eta_{LR}$ of the LR $O(N)$ models with respect to the bare
scaling dimension of the SR model.
Therefore, in the $d$-dimensional case with power--law exponent 
$d+\sigma$ one has
\begin{equation}
\label{AnomalousDimensionCorrection}
\eta_{LR}(d,\sigma) \equiv 2-\sigma + \delta\eta\,,
\end{equation}
in agreement with the definition employed in Ref.\,\cite{Fisher1972}.
As mentioned before, the non mean-field correction $\delta\eta$
has been conjectured to
vanish at all orders in perturbation theory\,\cite{Sak1973}.

We can summarize many of the efforts in the study of statistical mechanics models with weak LR couplings by identifying the
following crucial problems
\begin{itemize}
\item {\em (a)} the determination of $\sigma_*$ and the validation of Sak's scenario;
\item {\em (b)} the study of the universal behaviour in the region $d/2<\sigma\leq\sigma_*$;
\item {\em (c)} the answer to the question whether the critical properties
of a LR model in dimension $d$ with power-law exponent $d+\sigma$ 
can be, at least approximatively,
inferred from those of a SR model in an effective fractional dimension
$D_{\rm eff}$ (this result being exact in
the spherical $N\to\infty$ limit). 
\item {\em (d)} the understanding of the interplay between LR and SR interactions near $\sigma_*$.
\end{itemize}
Regarding the first point, we observe that the controversies
about the actual value of $\sigma_*$ have only partially a compelling 
quantitative motivation. Indeed, as an example,
for the two-dimensional Ising model with LR interactions, the
value $\sigma_*=7/4$ predicted by Sak is pretty close to
$\sigma_*=2$. Notice anyway that the value of
$\eta$ at $\sigma=7/4$ obtained in Ref.\,\cite{Blanchard2013} is 
$\eta = 0.332$, quantitatively different for the value $\eta=1/4$ predicted by Sak.
We think that the determination of $\sigma_*$ is rather a question of principle,
since it generally concerns how different momentum terms in the critical propagator contribute to the universal behaviour.
In particular, one may wonder how the LR ($p^\sigma$) term
renormalizes and how the SR term ($p^2$) in the propagator is dressed by the presence of LR
interactions. Both these questions constitute the essence of the last point in the list above.

In this work, we deal with the previous issues by first considering
the case of classical $O(N)$ models in dimension $d \ge 2$
for weak LR interactions, see Sections\,\ref{Sec1_Chap5}-\ref{SR_DR}.
The goal of these sections
is to review the FRG treatment for classical models and to set up the
ground for the study of the quantum case,
which is the subject of
Sections\,\ref{sec:2}-\ref{sec:4}. Our presentation is based on the formalism
developed and the results obtained in\,\cite{Defenu2015,Defenu2016,defenu17},
while the presentation follows the one in\,\cite{DefenuThesis}. Our main
objectives is to present such findings in an unified and compact way, providing
as well details on the derivations and new results on the topic of weak
LR spin systems.

An important motivation for the study of quantum spin models
with LR interactions is that several
recent technological applications generate 
a remarkable improvement of the experimental tools for the preparation and study
of atomic, molecular and optical systems 
and lead to the simulation 
of a wide range of quantum models\,\cite{Bloch2008}.
The realisation and investigation of equilibrium and dynamical behaviour of quantum systems in presence of LR interactions is one of the most exciting research topics\,\cite{Britton2012,Schauss2012,Aikawa2012,Lu2012,Yan2013,Firstenberg2013,Islam2013,Richerme2014,Jurcevic2014,Douglas2015,Schempp2015,Landig2015,Landig2015a,Hess2017,Jurcevic2017,Pagano2018,Hempel2018,Keeslin2019}.
Ising and/or $XY$ quantum spin chains with tunable LR interactions have been
realized in several ways, including $Be$ ions in a Penning trap\,\cite{Britton2012} or trapped ions coupled to 
motional degrees of freedom\,\cite{Islam2013,Richerme2014,Jurcevic2014}. 
The resulting interactions decay algebraically with 
the distance $r$ and the decay exponent can be experimentally tuned.
The possibility of controlling the range LR interactions 
in spin chains was also at the basis 
of the experimental simulation of the $1D$ Schwinger model with trapped ions
\cite{Martinez2016}. These exciting progress motivated a considerable
theoretical activity \cite{koffel2012,Schachenmayer2013,Knap2013,Hauke2013,Eisert2013,Gong2014,Damanik2014,Foss-Feig2015,Rajabpour2015,Gori2015,Cevolani2015,Hwang2015,Bertini2015,Santos2016,Gong2016,Humeniuk2016,Maghrebi2016,Kovacs2016,Gong2016a,Kuwahara2016,Buyskikh2016,Fey2016,celardo16,Viyuela2015,Bermudez2016,Cevolani16,Jakschke2017,Lepori2016,Cevolani18,igloi18,Lerose,blass18,defenu18,Lerose18,lerose19,Lepori2019,Fey2019}. Among the different quantum LR models studied, particular attention
has been devoted to LR Kitaev chains\,\cite{vodola14,Ares2015,lepori16,VanRegemortel2016,Vodola2016,lepori17,Dutta2017,Alecce2017,Defenu2019}.

The questions list above {\em (a)}$\cdots${\em (d)} equally applies
to weak LR quantum systems. Regarding the points {\em (a)} and {\em (b)},
 Sak's scenario applied to the quantum models implies that in the mean--field
approximation one has $\sigma_*=2$\,\cite{Dutta2001}, so that in the simplified
treatments, which neglect quantum fluctuations, one has the SR
universality class occurring at $\sigma>2$\,\cite{Knap2013}. The effect
of quantum fluctuations makes $\sigma_{*}<2$, and one can then study
$\sigma_*$ as a function of $d$ and $N$. Results for $\sigma_*$ and for the
critical exponents were derived
in\,\cite{defenu17} and are discussed in Sections\,\ref{sec:2}-\ref{sec:4}.
The point {\em (c)} requires, in practice, that one determines the equivalent
classical model at criticality\,\cite{Sachdev2011}, and, therefore, one needs to know the dynamical critical exponent $z$, which is computed in Section\,\ref{sec:4} extending the formalism presented in Sections\,\ref{Sec1_Chap5}-\ref{SR_DR}. The combined sequential discussion of the classical and of the quantum
cases aims at evidencing their common properties \& analogies (as well as their differences), and it is intended to clarify the point {\em (d)}, {\em i.e.}
how the LR terms behave under renormalization in presence of the SR one near
$\sigma_*$.

\section{FRG Approach to Classical Long-Range $O(N)$ models}
\label{Sec1_Chap5}
In this section we discuss the application of the FRG
approach\,\cite{Berges2002,Delamotte2012} to weak LR classical systems.
We are interested to critical properties and
 universal quantities, and, as usual, 
we replace the spin variables $\{\mathbf{S}_{i}\}$ 
with an $N$--component vector field $\boldsymbol{\phi}(x)$ in continuous space\,\cite{Mussardo2010}. 
We define the scale dependent effective action $\Gamma_{k}$ 
depending on the infrared cutoff scale $k$ and on the continuous field 
$\boldsymbol{\phi}$. When $k\rightarrow k_0$, where
$k_0$ is some ultraviolet scale, the effective action is equal to the
mean--field 
free energy of the system, while for $k\rightarrow 0$ it 
is equal to the exact free energy\,\cite{Delamotte2012}.
 
The FRG approach is based on the Wetterich flow equation for the effective action\,\cite{Wetterich1993}
\begin{align}
\label{Eq4}
\partial_{t}\Gamma_{k}=\frac{1}{2}{\rm Tr}\left[\frac{\partial_{t}R_{k}}{\Gamma^{(2)}+R_{k}}\right],
\end{align} 
where $k$ is a finite scale proportional to the
inverse system size ($k\propto L^{-1}$), $R_{k}$ is an infrared regulator function and
$t=\log\left(k/k_{0}\right)$ is the logarithmic scale, with $k_{0}$ the ultra-violet cutoff, which is proportional to the inverse lattice size. 
$\Gamma$ is the system effective action and represents the 
exact Ginzburg-Landau free
energy, whose second derivative with respect to the system internal variables $\Gamma^{(2)}$ gives the inverse system propagator.

An exact solution of Eq.\,\eqref{Eq4} is in general not possible. One shall rather find an approximate one by restricting the functional space of the possible effective action by making a proper ansatz. It is convenient to firstly consider a simplified parametrization
\begin{equation}
\label{EffectiveAction}
\Gamma_{k}[\phi]=\int d^{d}x \left\{Z_{k}\partial^{\frac{\sigma}{2}}_{\mu}
\phi_{i}\partial^{\frac{\sigma}{2}}_{\mu}\phi_{i}+U_{k}(\rho)\right\}\,,
\end{equation}
where $\rho=\frac{1}{2}\phi_{i}\phi_{i}$,
$\phi_{i}$ is the $i$--th component of $\boldsymbol{\phi}$ and the summation over repeated indexes is intended.
The fractional derivative $\partial^{\frac{\sigma}{2}}_{\mu}$ can be easily computed  
in Fourier space assuming periodic boundary conditions. Therefore, the second derivative of the action on Eq.\,\eqref{EffectiveAction} depends on $q^{\sigma}$ rather than on $q^{2}$ as in happens in the traditional case of SR interactions.
($q$ denoting here and in the following the modulus of $\vec{q}$).
The kinetic term $q^{\sigma}$ renormalizes only via the field independent coefficient $Z_{k}$, to which we refer as wave function renormalization.

The ansatz in Eq.\,\eqref{EffectiveAction} only includes the most relevant momentum term in the low energy behaviour of an $O(N)$ spin system and it can be deducted by 
the momentum dependence of the propagator at the mean-field level. Indeed, for $O(N)$ spins the mean-field propagator is given by the inverse coupling matrix $J_{ij}^{-1}$. The inverse of the coupling matrix is not analytically amenable for power-law decaying interactions, but a series expansion
in the momentum $q$ can be derived. Such series expansion
contains the non-analytic contribution $q^{\sigma}$, which, for $\sigma<2$, dominates the long wavelength behaviour. In principle, any modification in the low momentum limit of the propagation generated by the renormalization procedure should lead to the divergence of the wave-function renormalization $Z_{k}$ as it happens in the SR case.
For $\sigma\simeq 2$ the scaling of the non-analytic term approaches the SR analytic contribution, which is the first correction in the series expansion of the inverse coupling matrix $J_{ij}$. In the following the latter situation is represented by the ansatz in Eq.\,\eqref{EffectiveAction_LPA$''$}.

The renormalization of the effective potential $U_{k}(\rho)$ can be written in closed form in the FRG formalism by defining a regulator function $R_{k}(q)$, which introduces a momentum dependent mass term for the excitations with momentum $q<k$. In the case of non analytic corrections one may introduce the regulator function
$R_{k}(q)= Z_k (k^{\sigma}-q^{\sigma})\theta(k^{\sigma}-q^{\sigma})$,
which allows to the derivation of the flow equation for the effective potential.
In renormalization group calculations the critical point is identified by determining the fixed point of the evolution equations written in terms of dimensionless variables $\bar{U}_{*}(\bar{\rho})$ indicated by the bar superscript\,\cite{Codello2012}. Using dimensionless variables the evolution equation for the effective potential reads
\begin{equation}
\begin{split}
&\partial_t \bar{U}_{k}= -d\bar{U}_{k}(\bar{\rho})+(d-\sigma+\delta\eta)\bar{\rho}\,\bar{U}'_{k}(\bar{\rho})+\frac{\sigma}{2} c_d (N-1)\frac{1-\frac{\delta\eta}{d+\sigma}}{1+\bar{U}'_{k}(\bar{\rho})}
+\frac{\sigma}{2} c_d \frac{1-\frac{\delta\eta}{d+\sigma}}{1+\bar{U}'_{k}(\bar{\rho})+2\bar{\rho}\,\bar{U}''_{k}(\bar{\rho})}\,,
\label{EffectivePotentialLPAprimeEquation}
\end{split}
\end{equation}
with $c_d^{-1}=(4\pi)^{d/2} \Gamma\left(d/2+1 \right)$ 
and $\delta \eta = - \partial_t \log Z_k$, where $t=\log(k/k_0)$ is the RG time 
and $k_0$ is the ultraviolet scale. The quantity $\delta\eta$ indicates possible deviations from the mean-field momentum dependence of the propagator generated by the renormalization group procedure.
\subsection{Effective fractional dimension}
\label{eff_frac_dim}
As a preliminary computation we consider the case $Z_{k}=1$,
where by construction there are no anomalous dimension effects. 
With this simplified approach one can relate Eq.\,\eqref{EffectivePotentialLPAprimeEquation} to its SR counterpart as it appeared in Refs.\,\cite{Codello2012,Morris1994}. Indeed, Eq.\,\eqref{EffectivePotentialLPAprimeEquation} can be mapped to the evolution equation for a SR model in dimension $D_{{\rm eff}}$, which we refer to as the effective fractional dimension, whose expression reads
\begin{equation}
\label{DimensionaEquivalence}
D_{\rm eff}=\frac{2d}{\sigma}.
\end{equation}
Then, the effective potential for the LR system with parameters $(d,\sigma)$ is equal (apart for an unessential scaling factor) to the effective potential of a $D_{{\rm eff}}$-dimensional SR model. Also the correlation length and susceptibility exponents, $\nu_{LR}(d,\sigma)$ and $\gamma_{LR}(d,\sigma)$ respectively,
can be connected to their SR counterparts $\nu_{SR}(D_{\rm eff})$  
and $\gamma_{SR}(D_{\rm eff})$, using the relations%
\begin{equation}
\label{NuRelationLPA}
\nu_{LR}(d,\sigma)=\frac{2}{\sigma}\nu_{SR}(D_{\rm eff});\,\,\,\,\,
\gamma_{LR}(d,\sigma)=\gamma_{SR}(D_{\rm eff})\,.
\end{equation}
It is worth noting that the spherical and the Gaussian models, which  both have vanishing anomalous dimension in the SR limit, satisfy the relations in Eqs.\,\eqref{DimensionaEquivalence} and\,\eqref{NuRelationLPA} 
exactly\,\cite{Joyce1966}. 
In fact, the ansatz\,\eqref{EffectiveAction} becomes exact in the $N\to\infty$ limits, yielding exact results for the critical exponents.
In conclusion, we can see that the LR and SR  critical exponents
in, respectively, dimension $d$ and $D_{\rm eff}$ 
can be obtained from each other. Indeed, the two LR critical exponents $\nu_{LR}$  and $\gamma_{LR}$ can be derived from their SR equilvalents, using Eq.\,\eqref{NuRelationLPA}, 
and, then, the usual scaling relations can be used to obtain all the remaining critical exponents, including the anomalous dimension $\eta_{LR}=2-\sigma$. 
The equivalence between the fixed point structure between LR and SR models can also be seen using the spike plot technique
described in Refs.\,\cite{Codello2012,Morris1994}, see\,also Ref.\,\cite{DefenuThesis}
and Fig.\,1 of Ref.\,\cite{Defenu2015}.

From the analysis in Ref.\,\cite{DefenuThesis} it follows that  
fixing the dimension $d$ and increasing the values of $\sigma$ the LR interacting model presents a sequence of lower critical decay exponents $\sigma_{c,i}$ 
below novel universality classes appear in the phase diagram.
The origin of these multi-critical universalities is in close correspondence with the case of a SR system for fractional dimensions\,\cite{Codello2012}.
The lower critical decay exponent for the Wilson-FIsher universality, which describes the Ising model, is $\sigma_{c,2}=d/2$ and the LR model displays mean-field behavior for $\sigma<\sigma_{c,2}$\,\cite{Fisher1972}. 
The additional multi-critical universalities appear above the mean-field boundary for $\sigma>\sigma_{c,i}$, with $\sigma_{c,i}=d(i-1)/i$. In the limit $\sigma\to\sigma_{c,i}^{+}$ the correlation effects in the $i$-th universality class vanish and the ansatz in Eq.\,\eqref{EffectivePotentialLPAprimeEquation} becomes exact and it is the result for the existence of the $\sigma_{c,i}$.

The inclusion of a non trivial $Z_{k}$ value should allow to compute the anomalous dimension $\delta\eta$. Indeed, the evolution of the wave-function renormalization can be obtained via the relation
$\partial_tZ_{k}=\lim_{p\rightarrow 0}\frac{d}{dp^{\sigma}} \partial_t\Gamma_{k}^{(2)}(p,-p)$, which leads to the anomalous dimension according to the definition $\delta \eta = - \partial_t \log Z_k$.
However, the evolution of the inverse propagatopr $\Gamma_{k}^{(2)}(p,-p)$ remains analytic at small momenta $p\to 0$ also in presence of LR interactions, extending the $\delta\eta=0$ result also in the $Z_{k}\neq 1$ case. This result is in agreement with the investigation of Ref.\,\cite{Fisher1972} and it
leads to the Sak's result\,\cite{Sak1973} in the considered approximation
scheme, see Eq.\,\eqref{AnomalousDimensionCorrection}.

The introduction of renormalization effects in the kinetic sector of the model does produce a finite anomalous dimension in the SR model and, accordingly, an improved effective dimension expression can be introduced 
\begin{equation}
\label{EffectiveDimensionLPA$'$}
D_{\rm eff}'=\frac{[ 2-\eta_{SR}(D_{\rm eff}') ] d}{\sigma}\,,
\end{equation}
as it may be deducted by dimensional analysis\,\cite{Angelini2014} and with the arguments presented for the LR and SR Ising spin glasses in Ref.\,\cite{Banos2012}.
The expression in Eq.\,\eqref{EffectiveDimensionLPA$'$} is implicit and the critical exponent $\eta_{SR}$ as a function of the real parameter $d$ has to be computed
\cite{Codello2013,Katz1977,Holovatch1993,El-Showk2014,Cappelli2019}
in order to calculate $D_{\rm eff}'$. For consistency sake, we will use the $\eta_{SR}$ results in fractional dimension obtained by FRG.
\begin{figure}
\centering
\includegraphics[scale=.4]{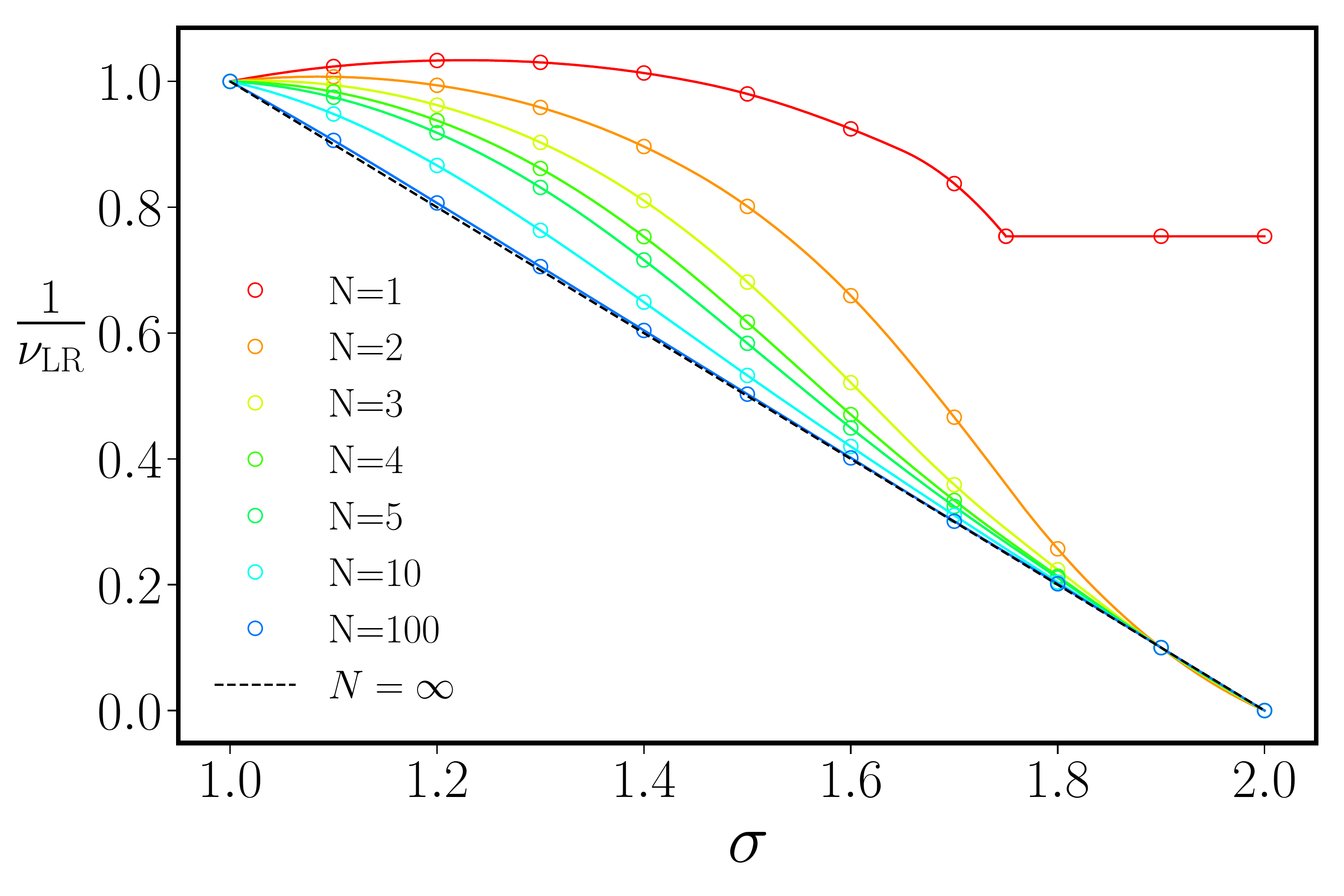}
\caption{
$y_{t}=1/\nu_{LR}$ exponent as a function of $\sigma$ in $d=2$ for some values of $N$ (from top: $N=1,2,3,4,5,10,100$).
The black dashed line is the analytical result obtained for the spherical model $N=\infty$.
}
\label{Fig1}
\end{figure}

The possibility to compute the value of $\nu_{{\rm SR}}$ for any real dimension $d$ also allows us to compute the correlation length exponent of the LR model at the present approximation level:
\begin{equation}
\label{NuRelationLPA2}
\nu_{LR}(d,\sigma)=\frac{2-\eta_{SR}(D_{\rm eff}')}{\sigma}\, \nu_{SR}(D_{\rm eff}')\,.\\
\end{equation}
The numerical curves for the inverse correlation length exponent  $y_{t}=1/\nu_{LR}$ are shown in Fig.\,\ref{Fig1} for several values of $N$. It is remarkable that for large $N$ the FRG approach the exact $N \to \infty$ limit result. Corresponding results for $d=3$, where the ansatz\,\eqref{AnomalousDimensionCorrection} is expected to be more reliable, have been presented in Ref.\,\cite{Defenu2015}.
The results for the universal quantities obtained by the ansatz in Eq.\,\eqref{EffectiveAction} become more reliable whenever the kinetic sector does not receive substantial contributions from the RG procedure and the leading term sticks to the mean field scaling. Therefore, our results become increasingly accurate when the anomalous dimension of the SR model in dimension $D_{{\rm eff}}$ is small.

\subsection{Effects of the short-range term}
The analysis presented in Sec.\,\ref{eff_frac_dim} suggests 
the validity of Sak's results for the value of $\sigma_*$. 
Nevertheless, due to the absence of the analytic term in the ansatz in Eq.\,\eqref{EffectiveAction}, 
the picture obtained in previous section is not reliable in the region
$\sigma\simeq\sigma_*$, where the two terms are competing. 
Such issue can be conveniently solved by the introduction of a more complete parametrisation, which accounts for the leading and first sub-leading term in the expansion of the mean-field propagator\,\cite{Defenu2015}:
\begin{equation}
\label{EffectiveAction_LPA$''$}
\Gamma_{k}[\phi]=\!\int\! d^{d}x\! \left\{\!Z_{\sigma,k}\partial^{\frac{\sigma}{2}}_{\mu}
\phi_{i}\partial^{\frac{\sigma}{2}}_{\mu}\phi_{i}\!+\!Z_{2,k}\partial_{\mu}\phi_{i}\partial_{\mu}\phi_{i}\!+\!U_{k}(\rho)\!\right\}\,.
\end{equation}
It is convenient to modify the regulator function employed in previous section in order not to introduce any bias
on the effect of the two competing terms. Indeed, in the region $\sigma\simeq\sigma_{*}$ one does not know what is the leading term responsible for the universal behaviour and one shall regularise both the terms as follows:
\begin{equation}
\label{Cutoff_Function_LPA$''$}
R_{k}(q)=Z_{\sigma,k} (k^{\sigma}-q^{\sigma})\theta(k^{\sigma}-q^{\sigma})+
Z_{2,k} (k^{2}-q^{2})\theta(k^{2}-q^{2})\,.
\end{equation}
It is immediate to check that both the ansatz in Eq.\,\eqref{EffectiveAction_LPA$''$} and regulator in Eq.\,\eqref{Cutoff_Function_LPA$''$} remain valid in the whole $\sigma\in(0,\infty)$ range. The regulator function in Eq.\,\eqref{Cutoff_Function_LPA$''$} acts symmetrically on both the LR $q^{\sigma}$ and the SR $q^{2}$ terms and it will introduce any bias in the competition between the terms.  Moreover, the choice in Eq.\,\eqref{Cutoff_Function_LPA$''$} consistently simplifies the calculation since it removes the momentum dependence from the low energy propagator.

The flow of the kinetic sector can be deducted by the renormalization group equation from the inverse propagator
\begin{align}
\label{Wave_Function_Flow_Definition}
\partial_t Z_{2}&=\frac{1}{2}\lim_{p\rightarrow 0}\frac{d^{2}}{dp^{2}} \partial_t \Gamma^{(2)}_{k}(p,-p)\\
\partial_t Z_{\sigma}&=\lim_{p\rightarrow 0}\frac{d}{dp^{\sigma}}\partial_t \Gamma^{(2)}_{k}(p,-p)\,,
\end{align}
which is obtained by the second derivative of the effective action evolution in Eq.\,\eqref{Eq4}. Using the ansatz in Eq.\,\eqref{EffectiveAction_LPA$''$} with the regulator function defined according to Eq.\,\eqref{Cutoff_Function_LPA$''$} one obtains the evolution equations
\begin{subequations}
\begin{equation}
\label{Dimensional_Z_Sigma_Flow_app}
\partial_t Z_{\sigma}=0\,,
\end{equation}
\begin{equation}
\label{Dimensional_Z_2_Flow_app}
\partial_t Z_{2}=
\!-\!\frac{\rho_{0}\,U''_{k}(\rho_{0})^{2}\,(\sigma Z_{\sigma}k^{\sigma}\!\!+\!2Z_{2})^{2}k^{d+2}}{(Z_{\sigma}k^{\sigma}\!+\!Z_{2}k^{2})^{2}(Z_{\sigma}k^{\sigma}\!+\!Z_{2}k^{2}\!+\!2\bar{\rho}_{0}U''_{k}(\rho_{0}))^{2}}\,,
\end{equation}
\begin{equation}
\label{Dimensional_Potential_Flow_app}
\partial_t U_{k}(\rho)=
\;\frac{Z_{2}k^{2}-\frac{\partial_t Z_{2}}{d+2}+\frac{\sigma}{2}Z_{\sigma}}{Z_{\sigma}k^{\sigma}+Z_{2}k^{2}+U'_{k}(\rho)+2\rho U''_{k}(\rho)}
+
(N-1)\frac{Z_{2}k^{2}-\frac{\partial_t Z_{2}}{d+2}+\frac{\sigma}{2}Z_{\sigma}}{Z_{\sigma}k^{\sigma}+Z_{2}k^{2}+U'_{k}(\rho)}\,.
\end{equation}
\end{subequations}
The next step in our calculation will be to rescale the physical quantities with an appropriate power of the scale $k$ in order to make the critical point a fixed point of the RG flow. The rescaling procedure defines the dimensionless couplings and it can be pursued according to two natural choices: the calculation in Sec.\,\ref{eff_frac_dim} have been performed in the assumption of a dimensionless $Z_{\sigma}$ coupling (which later absorbed into the field definition) we refer to this choice as to 
{\em LR-dimensions}. Another possible choice in the present case is to follow the path of SR $O(N)$ models and assume that the $Z_{2}$ wave-function is dimensionless and then absorbing it in the field. We refer to this formalism as {\em SR-dimensions}. In the latter framework the LR kinetic term is proportional to the
$J_{\sigma}=\frac{Z_{\sigma}}{Z_{2}}$ coupling.

The following table explains the two possible definitions of dimensionless couplings in details:
\begin{center}
\begin{tabular}{ccc}
{\em Quantity \quad \/} & {\em SR--dimensions \quad \/} & {\em LR--dimensions\/}\\
$q$ & $k\bar{q}$ & $k\bar{q}$ \\
$\rho$ & $k^{d-2} Z_{2}^{-1} \bar{\rho}$ & $k^{d-\sigma} Z_{\sigma}^{-1} \bar{\rho}$ \\
$U(\rho)$ & $k^{d}\bar{U}(\bar{\rho})$ &  $k^{d}\bar{U}(\bar{\rho})$\\
$Z_{2}$ & $\bar{Z}_{2}$ &  $k^{\sigma-2}\bar{Z}_{2}$\\
$Z_{\sigma}$ & $k^{2-\sigma}\bar{Z}_{\sigma}$ &  $\bar{Z}_{\sigma}$\\
\end{tabular}
\end{center}
Both definitions of dimensionless quantities shall yield the same physical results. 
In the SR-dimensions framework three equations are found, 
one for the potential $\bar{U}_{k}$, one for the LR coupling $J_{\sigma}$ and one for the anomalous dimension $\eta$. 
In the limit $J_{\sigma}\rightarrow 0$ this formalism reduce to the standard SR case. 
Conversely, in the LR-dimensions framework the anomalous dimension sticks to the mean-field value $\eta=2-\sigma$, while the  $J_{2}=Z_{2}/Z_{\sigma}$ remains finite only  for $\sigma<2$.

Due to the fact that the SR coupling $J_{2}$ remains finite in the whole $\sigma>0$ range, the SR-dimensions are the most suited to investigate the problem, without the need to switch between the two formalism in the region $\sigma\simeq\sigma_*$, where $Z_{2}$ starts diverging in the case of dominant SR interactions.
\section{SR-- and LR-- Dimensions}
\label{SR_DR}
As argued in the section above we will now use the ansatz in Eq.\,\eqref{EffectiveAction_LPA$''$} in the SR-dimensions formalism, where
the anomalous dimension is defined by
\begin{equation}
\eta_{2}=-\frac{1}{Z_{2}} \partial_t Z_{2} \,,
\end{equation}
as in the case of SR systems \cite{Delamotte2012}.
The additional LR contribution in proportional to the coupling $J_{\sigma}$ and the system is fully characterised by the following set of flow equations
\begin{subequations}
\begin{equation}
\label{J_Flow}
\partial_t \bar{J}_{\sigma} = (\sigma-2)\bar{J}_{\sigma}+\eta_{2}\bar{J}_{\sigma}\,,
\end{equation}
\begin{equation}
\eta_{2}=\frac{(2+\sigma\bar{J}_{\sigma})^{2}\bar{\rho}_0\bar{U}''_{k}(\bar{\rho}_0)^{2}}{(1+\bar{J}_{\sigma})^{2}(1+\bar{J}_{\sigma}+2\bar{\rho}_0\bar{U}''_{k}(\bar{\rho}_0))^{2}}\,,
\end{equation}
\begin{equation}
\begin{split}
\label{PotentialFlow_app}
&\partial_t \bar{U}_{k}(\bar{\rho})=-d\bar{U}_{k}(\bar{\rho})+(d-2+\eta_{2})\bar{\rho}\,\bar{U}'_{k}(\bar{\rho})\\
+&(N-1)\frac{1-\frac{\eta_{2}}{d+2}+\frac{\sigma}{2}\bar{J}_{\sigma}}{1+\bar{J}_{\sigma}+\bar{U}'_{k}(\bar{\rho})}
+\frac{1-\frac{\eta_{2}}{d+2}+\frac{\sigma}{2}\bar{J}_{\sigma}}{1+\bar{J}_{\sigma}+\bar{U}'_{k}(\bar{\rho})+2\bar{\rho}\,\bar{U}''_{k}(\bar{\rho})}\,.
\end{split}
\end{equation}
\end{subequations}
The only possible situations where the evolution of the LR coupling $J_{\sigma}$ in Eq.\,\eqref{J_Flow}  are either $\bar{J}_{\sigma}=0$ or $\eta=2-\sigma$. 

In the $\bar{J}_{\sigma}^{*}=0$ case the LR term vanishes and the universal behaviour of the system reduces to the case of SR interactions. When LR interactions are relevant, the only possibility to obtain a fixed point is  to have $\eta_{2}=2-\sigma$. As anticipated there is no necessity to switch between the SR- and the LR-dimensions formalisms even for relevant LR interactions. Indeed, the SR contibution $p^{2}$ in the propagator remains finite at the LR fixed point, but its scaling dimension renormalizes in order to match the scaling of the $p^{\sigma}$ term.

Qualitative aspects of the phase diagram can be investigated by expanding the effective potential around its non-trivial minimum
\begin{equation}
\label{Potential_Expansion_Around_Minimum_LPA$'$}
\bar{U}_{k}(\bar{\rho})=\frac{1}{2}\lambda_{k}(\bar{\rho}-\kappa_{k})^{2}\,.
\end{equation}
The $\beta$-functions for the two couplings $\lambda_{k}$ and $\rho_{k}$ are obtained by substituting the expression in Eq.\,\eqref{Potential_Expansion_Around_Minimum_LPA$'$} in the evolution for the effective potential, which together with the flow for $\bar{J}_{\sigma}$ can be used to introduce the following
closed set of equations:
\begin{subequations}
\label{RG_equations}
\begin{equation}
\label{jsig_eq}
\partial_{t}\bar{J}_{\sigma}=(\sigma-2)\bar{J}_{\sigma}+\eta_{2}\bar{J}_{\sigma}\,,
\end{equation}
\begin{equation}
\label{etaft}
\eta_{2}=\frac{(2+\sigma\bar{J}_{\sigma})^{2}\kappa_{k}\lambda_{k}^{2}}{(1+\bar{J}_{\sigma})^{2}(1+\bar{J}_{\sigma}+2\kappa_{k}\lambda_{k})^{2}}\,,
\end{equation}
\begin{equation}
\label{opportuna}
\partial_{t}\kappa_{k} =  \;   -(d-2+\eta_2 )\kappa_k + 3\frac{1-\frac{\eta_2 }{d+2}+\frac{\sigma}{2}\bar{J}_{\sigma}}{(1+\bar{J}_{\sigma}+2 \kappa_k  \lambda_k)^2}+\\
(N-1)\frac{1-\frac{\eta_2}{d+2}+\frac{ \sigma}{2}\bar{J}_{\sigma}}{(1+\bar{J}_{\sigma})^2}\,,
\end{equation}
\begin{equation}
\partial_{t} \lambda_{k} =(d-4+2\eta_{2})\lambda_{k} + 18 \lambda_k\frac{1-\frac{\eta_2 }{d+2}+\frac{\sigma}{2}\bar{J}_{\sigma}}{(1+\bar{J}_{\sigma}+2 \kappa_k  \lambda_k)^3}
+\\
 2 \lambda_k(N-1)\frac{1-\frac{\eta_2}{d+2}+\frac{ \sigma}{2}\bar{J}_{\sigma}}{(1+\bar{J}_{\sigma})^3}\,.
\end{equation}
\end{subequations}
The behaviour of $\eta_{2}$ obtained by Eq.\,\eqref{etaft} agrees with the one predicted by J. Sak. Indeed, the LR solution for the evolution equations disappears for $\sigma>\sigma_*$ and only the SR fixed point $\bar{J}_{\sigma}=0$ may exist. However, at $\sigma=\sigma_*$ another solution branches out from the SR one, with  $\eta_{2}=2-\sigma_{*}$ and $\bar{J}_{\sigma}=0$. For stronger LR interactions $\sigma<\sigma_{*}$ the LR coupling increases in order to satisfy the equation $\eta_{2}=2-\sigma$.
The increase in $\bar{J}_{\sigma,*}$ as $\sigma$ approaches the mean-field boundary $\sigma=d/2$ is rather steep. Indeed, the strength of LR interactions in the SR-dimensions formalism diverges for $\sigma=d/2$, where the LR correlated fixed point collides with the Gaussian one  and the SR term vanishes in the propagator.
We verified the above picture for several values of $d$ and $N$. 
Furthermore, the introduction of further couplings in the expansion of the effective potnetial in Eq.\,\eqref{Potential_Expansion_Around_Minimum_LPA$'$} does not spoil the above results nor the existence of the $\sigma_*$ boundary.

\begin{figure}
\centering
\includegraphics[scale=.33]{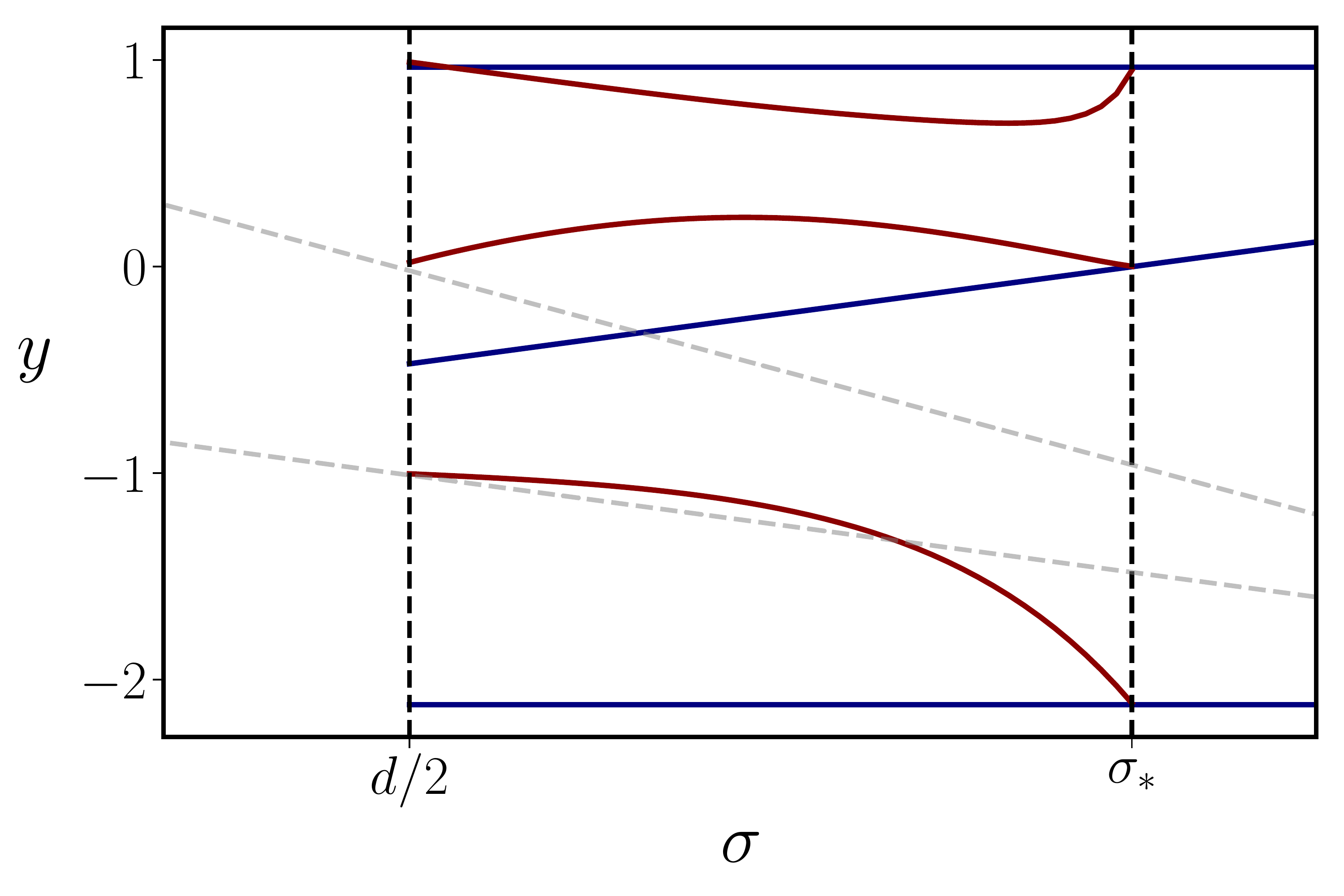}
\caption{The three eigenvalues $y$ of the RG stability matrix as a function of $\sigma$ for the LR  fixed point (red lines), the SR fixed point (blue lines) and the LR mean field fixed point (grey dashed lines) for the Ising model in $d=2$. The eigenvalues of the LR fixed points (both mean field and correlated) can be obtained either in the LR- or SR-dimensions scheme, leading to the same result. The blue lines for the SR fixed point have to be obtained in the SR-dimensions scheme.}
\label{Fig2}
\end{figure}

The relevance of each solution of the flow equations system in Eq.\,\eqref{RG_equations} can be deducted by the study of the perturbation spectrum.
For $\sigma>\sigma_*$ the SR fixed point only displays a single relevant direction (connected with the temperature) and no additional solution exists. 
Lowering  $\sigma$ towards the boundary $\sigma_*$ the eigenvalue of one of the two attractive direction decreases and finally hits zero at $\sigma_*$,
where the LR solution appears from the SR one.
Therefore, when LR are relevant, i.e. $\sigma<\sigma_*$, the repulsive directions of the SR fixed point become two and the LR solution is the leading one, since it has only one repulsive solution. Finally, in the mean-field limit $\sigma\to d/2$ the correlation effects disappear and the LR fixed point hits the Gaussian solution with $Z_{2}=0$. The perturbation spectrum of the flow equations system in Eq.\,\eqref{RG_equations} contains three eigenvalues, which are displayed in Fig.\,\ref{Fig2}. The eigenvalues of the LR Gaussian, LR and SR fixed points are presented by grey, red and blue lines respectively. Positive eigenvalues indicate attractive directions of the RG flows, the negative ones are repulsive.

The discussion above can be also reproduced in the-LR dimensions formalism. Indeed, using the wave function renormalisation $Z_{\sigma}$ and defining the
SR coupling $\bar{J}_{2}=\bar{Z}_{2}/Z_{\sigma}$ one obtains
\begin{subequations}
\label{set2}
\begin{equation}
\label{Dimensionless_Z_Sigma_Flow_app}
\partial_t Z_{\sigma}=0\,,
\end{equation}
\begin{equation}
\label{J_2_Flow_app}
\partial_t \bar{J}_{2}=(2-\sigma)\bar{J}_{2}-\frac{\bar{\rho}_0\,\bar{U}''_{k}(\bar{\rho}_0)^{2}(\sigma+2\bar{J}_{2})^{2}}{(1+\bar{J}_{2})^{2}(1+\bar{J}_{2}+2\kappa_{k}\bar{U}''_{k}(\bar{\rho}_0))^{2}}\,,
\end{equation}
\begin{equation}
\label{Dimensionless_Potential_Flow_app}
\begin{split}
\partial_t U_{k}(\rho)=&-d\bar{U}_{k}(\bar{\rho})+(d-\sigma)\bar{\rho}\,\bar{U}'_{k}(\bar{\rho})+\frac{\bar{J}_{2}-\frac{(2-\sigma)\bar{J}_{2}+\partial_t \bar{J}_{2}}{d+2}+\frac{\sigma}{2}}{1+\bar{J}_{2}+\bar{U}'_{k}(\bar{\rho})+2\bar{\rho} \,\bar{U}''_{k}(\bar{\rho})}\\
&+(N-1)\frac{\bar{J}_{2}-\frac{(2-\sigma)\bar{J}_{2}+\partial_t \bar{J}_{2}}{d+2}+\frac{\sigma}{2}}{1+\bar{J}_{2}+\bar{U}'_{k}(\bar{\rho})}\,.
\end{split}
\end{equation}
\end{subequations}
As expected the $\bar{J}_{2}\rightarrow 0$ limit of the latter system reproduces the results of ansatz in Eq.\,\eqref{EffectiveAction}, at least as long as $\sigma<\sigma_*$. Furthermore, one can compare Eqs.\,\eqref{set2} with the set in Eqs.\,\eqref{PotentialFlow_app}, the difference is of order $\bar{J}_{2}$ which remains very small for $\sigma<\sigma_{*}$  but diverges really fast close to the $\sigma_{*}$ boundary.

The previous analysis can then be repeated in the LR-dimensions formalism, but it will only be consistent as long as  $Z_{\sigma}$ remains finite, i.e. for $\sigma<\sigma_*$.
In the SR dominated regime the strength of the analytic term $p^{2}$ diverge, leading to the appearance of a finite non mean-field anomalous dimension.  
However, this divergence is not rescaled in the field when LR-dimensions are employed and then this scheme cannot work for $\sigma>\sigma_*$.

Therefore, we restrict the investigation of Eqs.\,\eqref{set2} to the $\sigma<\sigma_*$ regime, where LR-dimensions are well defined. The analysis will follow the same line as in the SR-dimension case, making use of the truncation in Eq.\,\eqref{Potential_Expansion_Around_Minimum_LPA$'$},  which leads to the evolution equations
\begin{subequations}
\label{lr_fl_set}
\begin{equation}
\begin{split}                     
\partial_{t}\bar{J}_{2}=(2-\sigma)\bar{J}_{2}+\frac{\kappa_{k}\lambda_{k}^{2}(\sigma+2\bar{J}_{2})^{2}}{(1+\bar{J}_{2})^{2}(1+\bar{J}_{2}+2\kappa_{k}\lambda_{k})^{2}}\,,
\end{split}
\end{equation}
\begin{equation}
\begin{split}\label{opportunabis1}
\partial_{t}\kappa_{k} =  \;   -(d-\sigma)\kappa_k + 3\frac{\bar{J}_{2}-\frac{(2-\sigma)\bar{J}_{2}+\partial_t \bar{J}_{2}}{d+2}+\frac{\sigma}{2}}{(1+\bar{J}_{2}+2 \kappa_k  \lambda_k)^2}+ (N-1)\frac{\bar{J}_{2}-\frac{(2-\sigma)\bar{J}_{2}+\partial_t \bar{J}_{2}}{d+2}+\frac{\sigma}{2}}{(1+\bar{J}_{2})^2}\,,
\end{split}
\end{equation}
\begin{equation}
\begin{split}\label{opportunabis2}
\partial_{t}\lambda_{k} =  \;   (d-2\sigma) \lambda_k +18 \lambda_k \frac{\bar{J}_{2}-\frac{(2-\sigma)\bar{J}_{2}+\partial_t \bar{J}_{2}}{d+2}+\frac{\sigma}{2}}{(1+\bar{J}_{2}+2 \kappa_k  \lambda_k)^3}+2 \lambda_k (N-1)\frac{\bar{J}_{2}-\frac{(2-\sigma)\bar{J}_{2}+\partial_t \bar{J}_{2}}{d+2}+\frac{\sigma}{2}}{(1+\bar{J}_{2})^3}\,.
\end{split}
\end{equation}
\end{subequations}
The Eqs.\,\eqref{lr_fl_set} only produce a single fixed point solution, i.e. the LR fixed point, and are only valid for  $\sigma<\sigma_*$, as it can be checked by investigating the divergence of $J_{2}$ in the $\sigma \to \sigma_*$ limit.
Despite this difference, the universal quantities calculated by Eqs.\,\eqref{lr_fl_set} coincide with the ones obtained by Eqs.\,\eqref{RG_equations}, as it should be. 

Before moving to the quantum case we would like to further comment on the difference between the correlation length exponent $\nu$ computed by the approximation in Eq\,\eqref{EffectiveAction_LPA$''$} and the one coming from the effective dimension approach, see Eq.\,\eqref{EffectiveDimensionLPA$'$}.
The comparison between the two approximation levels serves to estimate the error coming from the effective dimension approach. Such error remains smaller than $1\%$ also very close to the $\sigma_*$ boundary.
\section{The quantum rotor models}
\label{sec:2}
In this section two quantum LR models will be considered, i.e. 
the $d$-dimensional transverse field Ising model 
and the quantum rotor model. The two Hamiltonians read respectively
\begin{align}
\label{Eq1}
H_{{\rm I}}&=-\sum_{ij}\frac{J_{ij}}{2}\sigma_{i}^{z}\sigma_{j}^{z}-
h\sum_{i}\sigma_{i}^{x},\\
\label{Eq2}
H_{{\rm R}}&=-\sum_{ij}\frac{J_{ij}}{2}\hat{\boldsymbol{n}}_{i}\cdot\hat{\boldsymbol{n}}_{j}+\frac{\lambda}{2}\sum_{i}\mathcal{L}_{i}^{2}, 
\end{align}
where $\sigma^{\mu}$ is the Pauli matrix in the $\mu$-direction, the $\hat{\boldsymbol{n}}_i$ are $N$ components unit length vector 
operators ($\hat{\boldsymbol{n}}_{i}^{2}=1$), $\lambda$ and $h$ are real constants. 
The operator $\mathcal{L}$ is invariant
under $O(N)$ rotations and it is formed from the asymmetric rotor space angular momentum tensor\,\cite{Sachdev2011}. Both the Ising spins and the quantum rotors are placed on the sites of a $d$-dimensional hypercubic lattice labeled by the indexes $i,j$. As in the previous sections, the interaction matrix decays as a power-law with the inter-site distance $r_{ij}$:
\begin{align}
\label{Eq3}
J_{ij}=\frac{J}{r_{ij}^{d+\sigma}},
\end{align}
with $J$ a positive constant.

At finite temperature the $O(N)$ quantum rotor models belong to the same universality class as the classical $O(N)$ field theories, while in the $T\to0$ limit quantum fluctuations become relevant altering the universal behavior\,\cite{Sachdev2011}.

Several many body quantum systems display the same universal behaviour of $O(N)$ quantum rotors. Especially, the quantum spherical model, the Heisenberg model with anti-ferromagnetic exchange and the Bose-Hubbard model belong respectively to the $N=\infty,3,2$ universalities. Moreover, the $N=2$ quantum rotors with LR interactions describes the critical behaviour of the isotrpic-nematic transitions observed in stripe forming systems with competing interactions\,\cite{Coto2017}. Finally, the $N=1$ rotor model displays $\mathbb{Z}_{2}$ symmetry and, then, it shares the same universal behaviour of the transverse Ising model described by the Hamiltonian in Eq.\,\eqref{Eq1}.

Once again, we only consider the $\sigma>0$ case in order to
have a well defined thermodynamics\,\cite{Campa2009,Campa2014}. At zero temperature both the Ising model and the general rotor models, which also encompass the Ising universality class in the $N=1$ case, do have in dimension $d>1$ a quantum phase transition (QPT)
as a function of the control parameters $h$ and $\lambda$ respectively.
On the other hand, the Mermin Wagner Theorem (MWT) forbids spontaneous symmetry breaking (SSB) for continuous symmetries $N\geq2$ in $d=1$, unless LR interactions are relevant for $\sigma<\sigma_{*}$. Also in the quantum case
we denote by $\sigma_{*}$ the boundary value separating the SR universality at $\sigma\geq\sigma_{*}$ from the LR one $\sigma<\sigma_{*}$.
As it is well known, the case $d=1$, $N=1$ features instead a
QPT also in the SR limit.

The classical phase transition (CPT), which arises at finite temperatures $T>0$, displays the same universal behaviour as classical LR $O(N)$ spin systems. Therefore, results for the finite 
temperature transition
behavior presented in the previous sections will be used in the following.


The solution of Eq.\,\eqref{Eq4} shall be found in the restricted functional space spanned by a finite number of functions and/or couplings. In
the field theoretical formalism one can use the Trotter decomposition, 
to show that the Hamiltonians defined in Eqs.\,\eqref{Eq1} 
and\,\eqref{Eq2} belong to the same universality class of their classical equivalent at finite temperature in $d+1$ dimension. The additional dimension is necessary to account for the propagation of quantum fluctuations, which may also depend on time. Therefore, the LR interactions in the effective model are present only in the $d$ spatial directions and SR in the remaining ``time'' dimension\,\cite{Dutta2001}. 

As a consequence, the generalisation of the ansatz in Eq.\,\eqref{EffectiveAction_LPA$''$} for the effective action of an $O(N)$ quantum rotor model reads
\begin{align}
\label{Eq5}
\Gamma_{k}=\int\,d\tau\,\int\,d^{d}x\{&K_{k}\partial_{\tau}\varphi_{i}\partial_{\tau}\varphi_{i}-Z_{k}\varphi_{i}\Delta^{\frac{\sigma}{2}}\varphi_{i}
-Z_{2,k}\varphi_{i}\Delta\varphi_{i}+U_{k}(\rho)\}
\end{align}
where $d$-dimensional in the spatial directions is represented by $\Delta$, while $\tau$ is the ``Trotter"/imaginary time direction. Analogously to the classical case, the vector field $\varphi_{i}(x)$ has $N$ components labeled by the index ($i \in \{1,\cdots,N\}$) and its squared modulus $\rho\equiv \sum\varphi_{i}^{2}/2$ gives the order parameter. In Eq.\,\eqref{Eq5} the summation over repeated indexes is intended.

The approximation introduced in ansatz\,\eqref{Eq5} is sufficient to give a semi-quantitative description of the universal behavior of the Hamiltonians in Eqs.\,\eqref{Eq2} and \eqref{Eq1} with numerical results for the critical exponents which agrees within 5\% with exact numerical computations (where present). When the SR kinetic term $\varphi_{i}\Delta\varphi_{i}$ becomes relevant one recovers the isotropic SR quantum $O(N)$ field theories, where the quantum to classical correspondence $d\to d+1$ holds exactly. In the $\sigma\simeq 2$ region the scaling dimension of the LR and SR kinetic terms are similar and their competition leads to the appearance of the $\sigma_{*}\leq2$ threshold.

At the critical point the appearance of scaling relations connects the frequency and momentum dependencies in the propagator leading to the dynamical critical exponent $z$, defined by $\omega\propto q^{z}$. Moreover, at large wavelength the bare momentum dependence of the propagator is altered by the presence of critical fluctuations, leading to the asymptotic relation
\begin{align}\label{lim}
\lim_{q\to 0}G(q)\propto q^{2-\eta},
\end{align}
which defines the anomalous dimension $\eta$. It is worth noting that in Eq.\,\eqref{lim} the critical exponent $\eta$ has been defined as the displacement of the long wavelength behaviour of the spatial propagator with respect to the SR mean field 
behavior $q^{2}$, in analogy with the classical case. In order to fully characterise the quantum critical point it is necessary to introduce a further critical exponent, apart from $z$ and $\eta$. In FRG applications it is customary to consider the correlation length critical exponent $\nu$ defined as
\begin{align}
\label{Eq7}
\xi\propto \left(\lambda-\lambda_{c}\right)^{-\nu}
\end{align}
where the coupling $\lambda$ introduced in the Hamiltonian in Eq.\,\eqref{Eq2} controls quantum fluctuations and the critical point appears at $\lambda=\lambda_{c}$. 
The three critical exponents $(z,\eta,\nu)$ uniquely identify the universal behaviour of any QCP, while further exponents can be obtained by the scaling relations\,\cite{Sachdev2011}. 

\subsection{Local potential approximation}

Before dealing with the full set of flow equations, it is convenient to construct a simpler description of the ansatz in Eq.\,\eqref{Eq5}. Therefore, we impose the additional constraints
\begin{align}
&Z_{k}=K_{k}=1,\nonumber\\
&Z_{2,k}=0,
\end{align}
and we are left with a single flow equation which describes the FRG evolution of the local potential $U_{k}(\rho)$, this approximation is ofter referred as local potential approximation (LPA). Following the same procedure outlined in the previous sections for the classical case, one obtains
\begin{align}
\label{Eq9}
\partial_{t}\bar{U}_{k}(\rho)&=(d+z)\,\bar{U}_{k}(\bar{\rho})-(d+z-\sigma)\,\bar{\rho}\,\bar{U}^{(1)}(\bar{\rho})-\frac{1}{1+\bar{U}^{(1)}(\bar{\rho})+2\,\bar{\rho}\bar{U}^{(2)}(\bar{\rho})}-\frac{N-1}{1+\bar{U}^{(1)}(\bar{\rho})}.
\end{align}
Here and in the following the rescaled quantities are indicated by the bar superscript. 
The definition of rescaled quantities can be obtained by generalising  the one indicated in the table following Eq.\,\eqref{Dimensional_Potential_Flow_app} to the quantum case. This definition ensures that the effective action remains unaltered by the renormalization procedure $\Gamma=\bar{\Gamma}$
\begin{align}
U(\rho)&=k^{D_{U}}\bar{U}(\bar{\rho})\\
\varphi&=k^{D_{\varphi}}\bar{\varphi},
\end{align}
with
\begin{align}
D_{U}&=d+z\\
D_{\varphi}&=d+z-\sigma.
\end{align}
The absence of renormalization for kinetic terms in the LPA approximation leads to vanishing anomalous dimension effect both in the frequency and momentum sectors. 
According to this argument the dynamical critical exponent and the spatial anomalous dimension should stick to the mean field result
\begin{align}
\eta&=2-\sigma,\label{eta_mf}\\
z&=\frac{\sigma}{2}.\label{z_mf}
\end{align}
A careful inspection of Eq.\,\eqref{Eq9} reveals that the flow for the effective potential in the quantum case at LPA level can be mapped to the one of the classical case \,\cite{Codello2013,Codello2015} 
in the effective fractional dimension 
\begin{align}
d_{{\rm SR}}=\frac{2(d+z)}{\sigma}.
\end{align}
The latter results has been obtained in LPA and does not account for the renormalization of the kinetic sector. Therefore,  it is consistent to employ the mean-field result for the exponent $z$ in the expression for the effective dimension above, leading to the explicit result
\begin{align}
d_{{\rm SR}}'=\frac{2d}{\sigma}+1,
\label{Eq17}
\end{align}
which generalises the well known quantum to classical correspondence to the LR case. 
Similar effective dimension relations have already been discussed in the exactly solvable case of the spherical model\,\cite{Joyce1966,Vojta1996} and for statistical mechanics models defined on diluted lattices\,\cite{Dutta2003}, where the 
random disorder is introduced to simulate LR effects. It is fundamental to note that while the traditional quantum to classical correspondence is an exact relation the use of effective dimension approaches in LR systems is only approximate, see the discussion in Ref.\,\cite{Defenu2015} as well as Refs.\,\cite{Behan2017,Connor2019}.

Since the effective dimension relations become exact in the vanishing anomalous dimension limit, they can be used to recover the exact results for the upper and lower critical dimension for the existence of a non trivial quantum phase transition. In the SR limit the upper critical dimension is $d_{{\rm uc}}=4$, leading to the result
\begin{align}
\label{Eq19}
d_{{\rm uc}}=\frac{3}{2}\sigma, 
\end{align}
in the LR case, in agreement with the prediction obtained by relevance arguments in Ref.\,\cite{Dutta2001}. Above the upper critical dimension $d>d_{uc}$ the LR model belong to the mean field universality with the critical exponents given in 
Eqs.\,\eqref{eta_mf} and\,\eqref{z_mf} and the correlation length exponent
\begin{align}
\nu&=\sigma^{-1}.\label{nu_mf}
\end{align}

The anomalous dimension corrections to the momentum and frequency terms in the critical propagator also vanish in the $N\geq2$ case for SR interactions at a lower critical dimension $d_{{\rm lc}}$, whose existence is ensured by the MWT. 
In the LR models the expression for the lower critical dimension of continuous symmetries, obtained by the effective dimension approach, reads
\begin{align}
\label{Eq20}
d_{{\rm lc}}=\frac{\sigma}{2}.
\end{align}
While the validity of the result in Eq.\,\eqref{Eq19} is independent on $N$, 
the lower critical dimension expression in Eq.\,\eqref{Eq20} is a consequence of the MWT and it applies only to continuous symmetries $N\geq 2$.

The most fascinating feature of the LPA picture is its simplicity. 
The solution of Eq.\,\eqref{Eq9} reveals the existence of two fixed points
for the effective potential. The mean field solution yields the critical exponents in Eqs.\, 
\eqref{eta_mf}, \eqref{z_mf} and \eqref{nu_mf}.
The second solution is interacting with finite renormalized mass 
$\bar{U}^{(1)}(0)\neq 0$ and describes the Wilson-Fisher universality.

The Gaussian fixed point is attractive in all but one directions
in the theory space for $d\geq d_{{\rm uc}}$, leading to
mean field universal behaviour for the quantum critical point. 
On the other hand, for systems in dimension $d<d_{{\rm uc}}$ the interacting fixed point solution dominates the FRG trajectories and the universality class is the Wilson-Fisher one with finite anomalous dimensions, which can be only approximatively studied. The case of continuous symmetries $N\geq2$ also presents a region $d\leq d_{{\rm lc}}$ where no SSB is possible and no finite order parameter is possible. 

Close to the relevance threshold for LR interactions $\sigma\simeq 2$ the competition between LR and SR terms in the propagator leads to an increasing importance of the anomalous dimension corrections. However, in LPA  
anomalous dimension effects are discarded and one retrieves the mean-field result $\sigma_{*}=2$\,\cite{Dutta2001}. However,  the correlated result for the boundary between LR and SR universalities should be given, also in the quantum case,
by the Sak's expression $\sigma_{*}=2-\eta_{{\rm SR}}$ 
\cite{Sak1973}. Since for irrelevant LR interactions the effective field theory model recovers isotropy in time and space the anomalous dimension $\eta_{{\rm SR}}$ of the SR model is the sameone of the classical model in $d+1$ dimension. In order 
to investigate this effect it is necessary to study the flow equations for all the couplings defined in Eq.\,\eqref{Eq5}. 

\subsection{Anomalous dimension effects}

In the approximation defined by the ansatz in Eq.\,\eqref{Eq5} the evolution of the coefficients $Z_{k}$, $K_{k}$ and $Z_{2,k}$ under renormalization group transformation fully describe the kinetic sector of the LR model. 
The wave-function $Z_{k}$ evolves according to the flow equation
\begin{align}
\label{Eq21}
\partial_{t}Z_{k}=\left(2-\sigma-\eta\right)Z_{k},
\end{align}
where the anomalous dimension $\eta$ is defined with respect to the SR term
\begin{align}
\eta=\frac{\partial_{t}Z_{2,k}}{Z_{2,k}}.
\end{align}
Given that the flow equations must vanish at the fixed points only two solutions are possible for the anomalous dimension according to Eq.\,\eqref{Eq21}, either
\begin{align}
\label{Eq22}
\eta=2-\sigma
\end{align}
or
\begin{align}
\label{Eq23}
\lim_{k\to 0}Z_{k}=0.
\end{align}
In the first case, when Eq.\,\eqref{Eq22} is fulfilled, the scaling of the SR term is renormalized to equal match the one of the LR term. Conversely, when Eq.\,\eqref{Eq22} is vakid LR interactions are irrelevant
and the system belong to the SR universality, consistently with the Sak's result.

The boundary value $\sigma_{*}$ can be identified by imposing that Eq.\,\eqref{Eq21} 
is fulfilled by the SR anomalous dimension value $\eta_{{\rm SR}}$, leading to the Sak's expression $2-\sigma_{*}-\eta_{{\rm SR}}=0$
\begin{align}
\label{Eq24}
\sigma_{*}=2-\eta_{{\rm SR}},
\end{align}
in perfect analogy with the classical case.
At $\sigma\geq\sigma_{*}$ isotropy in the spatial and time directions is recovered yielding the SR result $z=1$ and the traditional quantum to classical correspondence between the QCP and the thermal CPT is obeyed\,\cite{Mussardo2010,Sachdev2011}.

Merging the result for the correlated $\sigma_{*}$ value obtained in the present Section, with the upper critical dimension expression given in Eq.\,\eqref{Eq19} it is possible to depict the full phase diagram of the quantum LR rotor models in the $(d,\sigma)$ plane. Fig.\,\ref{Fig3} reports such phase diagram for the Ising universality ($N=1$) in $d=1$. The red solid line in Fig.\,\ref{Fig3} represents the
$\sigma_{*}$ boundary according to Eq.\,\eqref{Eq24} with $\eta_{{\rm SR}}$ given by the FRG result at the same approximation level. The relevant regime for LR interactions is given by the yellow shaded area in Fig.\,\ref{Fig3}, while the LR mean field universality appears in the green shaded area.

The phase diagram for continuous symmetries is equivalent to the one depicted in Fig.\,\ref{Fig3}, except that, due to the MWT, one has $\sigma_{*}=2$ 
both at $d=d_{{\rm uc}}$ and at $d=d_{{\rm lc}}$. In addition, for $N\ge2$ case only the trivial phase exists at $d\leq d_{{\rm lc}}$.
\begin{figure}[!h]
    \centering
            \includegraphics[width=.7\textwidth]{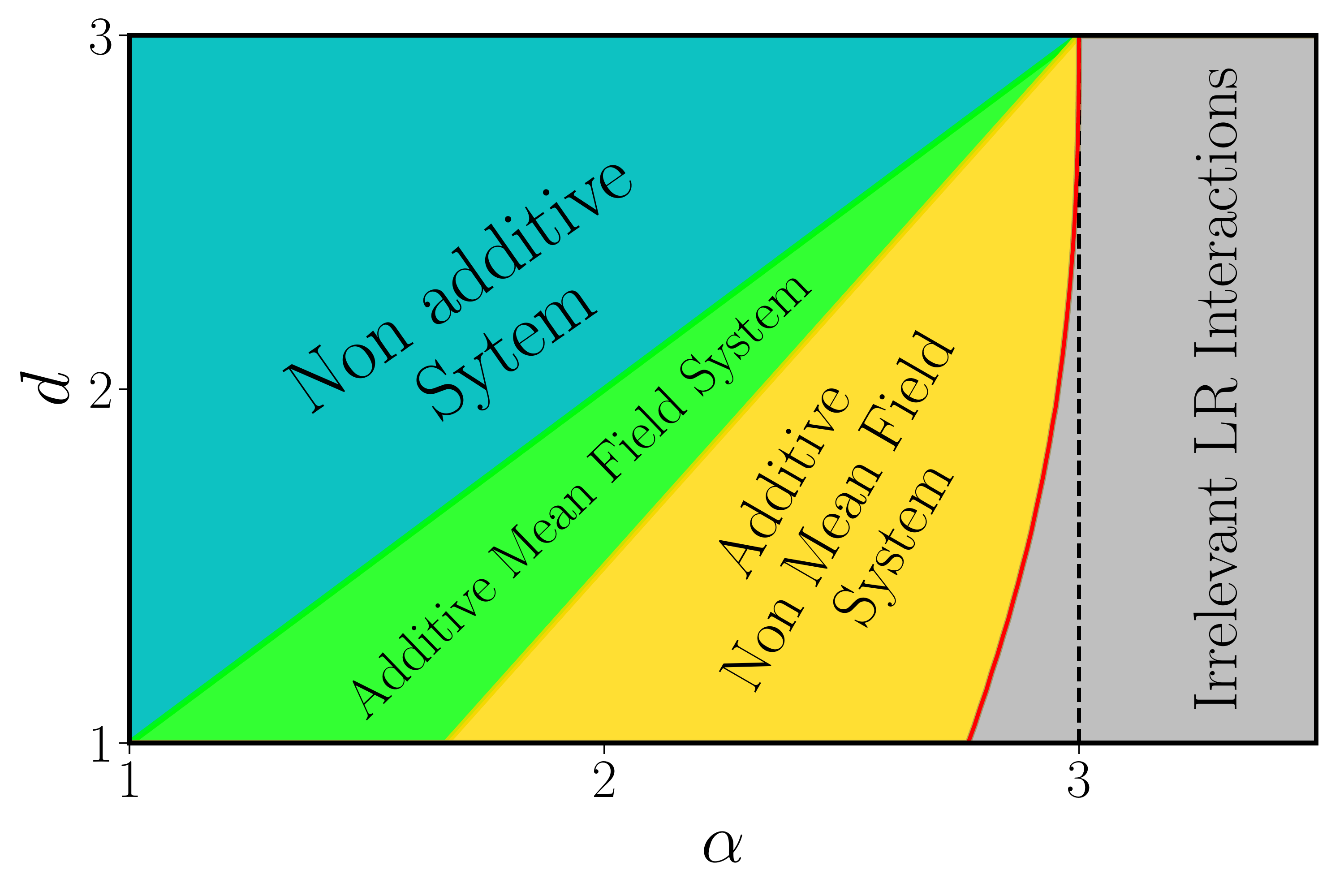}
    \caption{\label{Fig3} Phase diagram of the quantum Ising chain in a transverse field ($N=1$ and $d=1$) with LR couplings.
The green shaded area represents the mean field validity region, 
while the gold shaded area is the region with peculiar LR critical exponents. 
The mean field boundary between SR and LR universalities $\sigma_{*}=2$ is represented by a black dashed line, while the renormalized boundary is shown as a red solid line. Finally, the region with diverging interaction energy and ill defined thermodynamic functions is given as a blue background.}
    \label{Fig3}
\end{figure}

\subsection{The critical exponents}\label{sec:4}

The similarity between the phase diagram discussed in Sec.\,\ref{sec:4} and the one obtained for classical systems in Sec.\,\ref{Sec1_Chap5}, see also Refs.\,\cite{Angelini2014,Defenu2015} is evident, due to analogy between the $\sigma_{*}$ expressions.
In both cases, for $\sigma<\sigma_{*}$ the QCP of the theory is strongly interacting and the computation of universal quantities can only be pursued in approximate fashion.

The application of the FRG approach to quantum systems closely follows the path already explored in the case of anisotropic classical systems\,\cite{Defenu2016}. In order to simplify the computation, we consider the ansatz in Eq.\,\eqref{Eq5} for $Z_{2,k}=0$. This additional approximation does not introduce any further correction to the determination of the critical exponents, since the analytic momentum term only becomes relevant for $\sigma\simeq\sigma_{*}$  and, even there, it has been shown
in Ref.\cite{Defenu2015,Defenu2016} not to substantially influence the numerical values of the critical exponents.

The evolution equation for $\bar{U}_{k}$ in the $Z_{2,k}=0$ assumption reads
\begin{equation}
\label{Eq25}
\begin{split}
\partial_t \bar{U}_{k}= &(d+z)\bar{U}_{k}(\bar{\rho})-(d+z-\sigma)\bar{\rho}\,\bar{U}'_{k}(\bar{\rho})
- \frac{\sigma}{2}(N-1)\frac{1-\frac{\eta_{\tau} z}{3\sigma+2d}}{1+\bar{U}'_{k}(\bar{\rho})}
-\frac{\sigma}{2}\frac{1-\frac{\eta_{\tau} z}{3\sigma+2d}}{1+\bar{U}'_{k}(\bar{\rho})+2\bar{\rho}\,\bar{U}''_{k}(\bar{\rho})},
\end{split}
\end{equation} 
with $\eta_{\tau}$ being the interaction correction
with respect to the mean-field frequency dependence of the 
critical propagator
\begin{align}
\label{Eq26}
\lim_{\omega\to 0}G(\omega,1)^{-1}\propto \omega^{2-\eta_{\tau}}.
\end{align}
Analogously to the momentum anomalous dimension the correction $\eta_{\tau}$ can be obtained by the flow of $K_{k}$ 
\begin{align}
\label{Eq27}
\eta_{\tau}=\frac{f(\bar{\rho}_{0},\bar{U}^{(2)}(\bar{\rho}_{0}))(3\sigma+2 d)}{d+(3\sigma+d)(1+f(\bar{\rho}_{0},\bar{U}^{(2)}(\bar{\rho}_{0})))}.
\end{align}
The expression $f(\bar{\rho}_{0},\bar{U}^{(2)}(\bar{\rho}_{0}))$ 
is the same one found for the anomalous dimension of SR  
$O(N)$ models, see Ref.\,\cite{Codello2013}.
\begin{equation}
\label{Eta_SR}
\eta_{{\rm SR}}=f(\bar{\rho}_{0},\bar{U}^{(2)}(\bar{\rho}_{0}))=\frac{4\bar{\rho}_{0}\bar{U}^{(2)}(\bar{\rho}_{0})^{2}}{(1+2\bar{\rho}_{0}\bar{U}^{(2)}(\bar{\rho}_{0}))^{2}}.
\end{equation}
The frequency anomalous dimension $\eta_{\tau}$ is necessary to obtain the dynamical critical exponent $z$. Indeed, generalising Eq.\,\eqref{Eq26} to the frequency sector one obtains 
\begin{align}
z=\frac{\sigma}{2-\eta_{\tau}},
\end{align}
which reproduces the mean-field expression in Eq.\,\eqref{z_mf}.

\begin{figure}[!h]
\centering
\includegraphics[width=.75\textwidth]{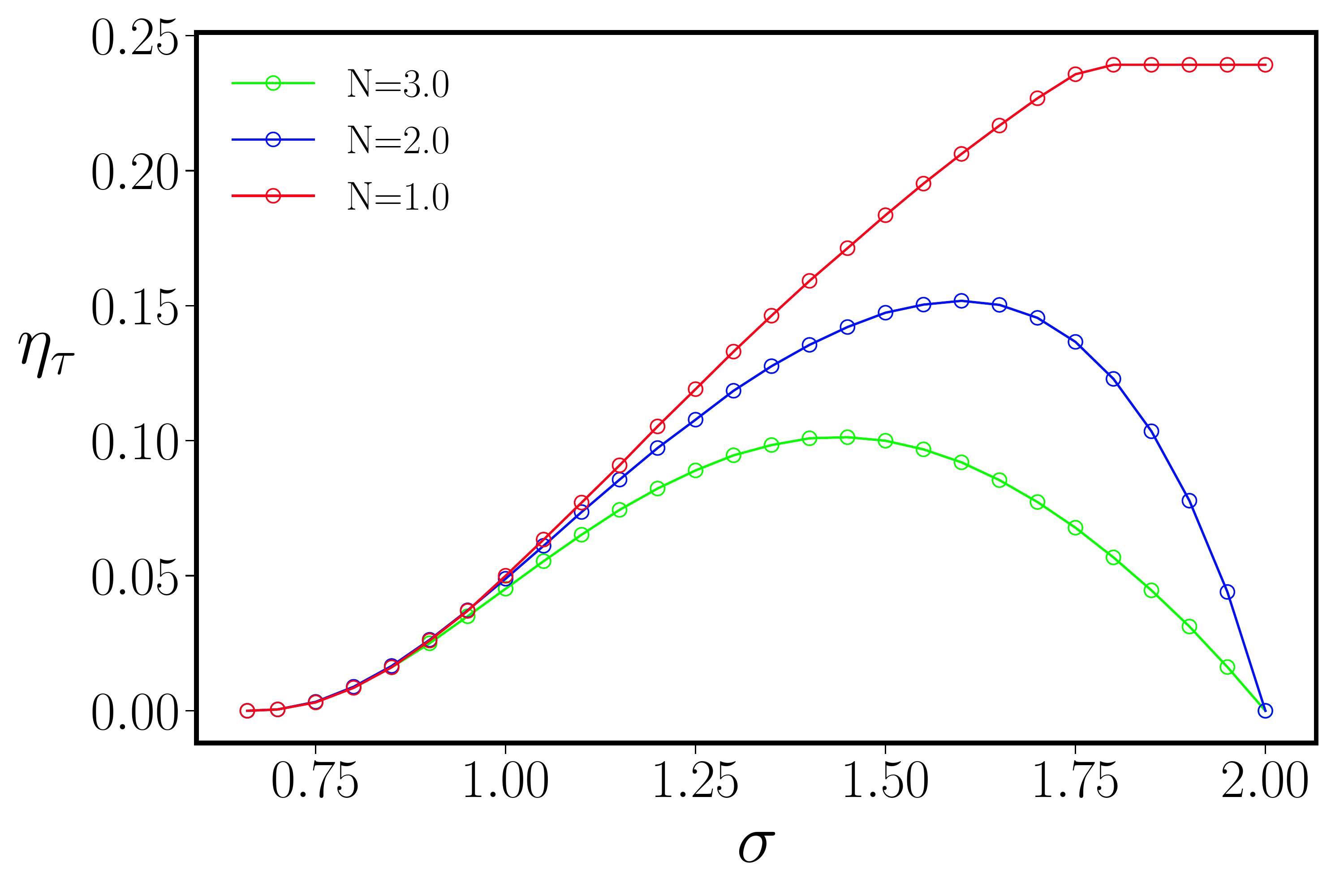}
  \caption{\label{Fig4} The anomalous dimension of the low frequency contribution to the critical propagator as a function of $\sigma$ in $d=1$. The green, blue and red 
curves are for $N=1,2,3$ respectively. 
The (blue and red) curves vanish in the SR limit $\sigma\to2$ due to the MWT.}
\end{figure}

In Fig.\,\ref{Fig4} the anomalous dimension 
$\eta_{\tau}$ as a function of the decay exponent $\sigma$ is shown in the $d=1$ case
(similar results can be obtained in $d=2$ \cite{Defenu2015}).
The data have been obtained solving 
the expression for the fixed point effective potential, 
Eq.\,\eqref{Eq25} with the l.h.s posed to zero, 
and Eq.\,\eqref{Eq27} in a self-consistent cycle.

The critical exponent $\eta_{\tau}$ vanishes below the mean field boundary $\sigma<\frac{2}{3}d$, while it becomes a non trivial curve in 
the non Gaussian region. It should be noted that correlation effects 
both increase the value of the anomalous dimension $\eta_{\tau}$ and of the dynamical critical exponent $z$\,\cite{Defenu2016}
with respect to the mean field prediction. However in the continuos symmetry 
cases $N=2,3$, for $d=1$, due to the MWT the curves bend 
down at some finite value $\sigma$ and the anomalous dimension $\eta_{\tau}$ 
vanishes at $\sigma_{*}=2$ in agreement with the MWT. 
On the other hand in the $d=2$ case and for all dimensions in the Ising 
case the curves are monotonically increasing and they met the SR line
at some value $\sigma_{*}<2$.

The present FRG approach has the indisputable advantage to capture, within a single calculation a consistent picture for the SSB transition in the whole $d$, $\sigma$ and $N$ parameter space, as a down side the topological 
phase transition which shall occur for $\sigma=2$ in $d=1$ can only be addressed within the present formalism\,\cite{DefenuBKT}. In passing we note that the MC results 
for a quantum XY model coupled 
to an anomalous Bosonic bath\,\cite{Sperstad2012} seem
to provide a value $\sigma_{*}=2-\eta$ with $\eta=\frac{1}{4}$ 
even for the $N=2$, which is consistent with the BKT scenario [see
\cite{Miguel} for a recent numerical study of the dilute LR
XY model]. 

The correlation length critical exponent in the non trivial region can be derived by studying the stability of the fixed point solution upon small perturbation in the parameter space. Then, we substitute in the RG evolution of the effective potential in Eq.\,\eqref{Eq25} the expression $\bar{U}(\bar{\rho})=\bar{U}_{*}(\bar{\rho})
+u(\bar{\rho})e^{y_{t}\,t}$, where  $\bar{U}_{*}(\bar{\rho})$ satisfies Eq.\,\eqref{Eq25} when the l.h.s. vanishes. 
\begin{figure}[!h]
    \centering
    \begin{subfigure}[b]{0.49\textwidth}
                \centering
            \includegraphics[width=\textwidth]{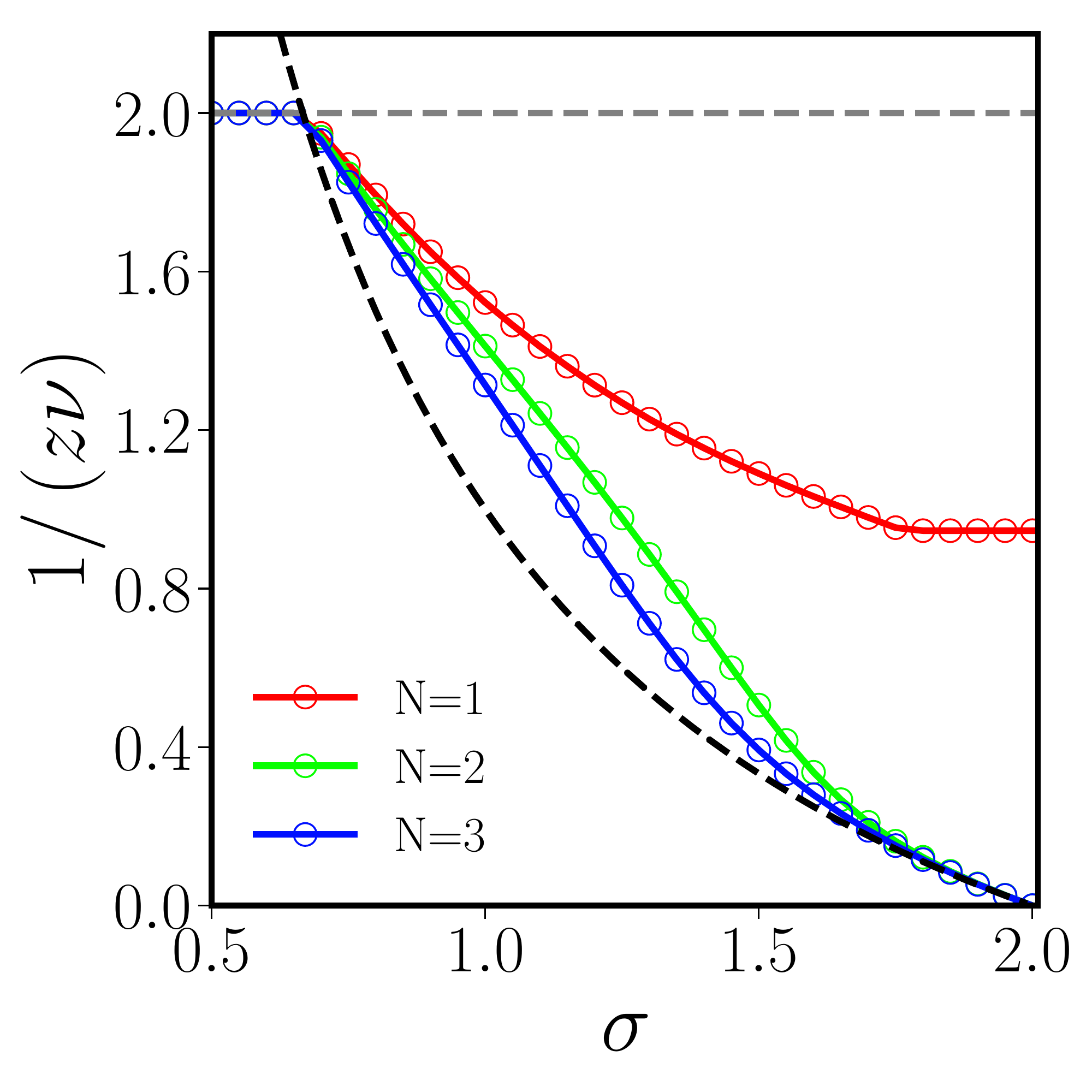}
    \caption{}
    \label{Fig5a}
    \end{subfigure}
\begin{subfigure}[b]{0.49\textwidth}
            \centering
            \includegraphics[width=\textwidth]{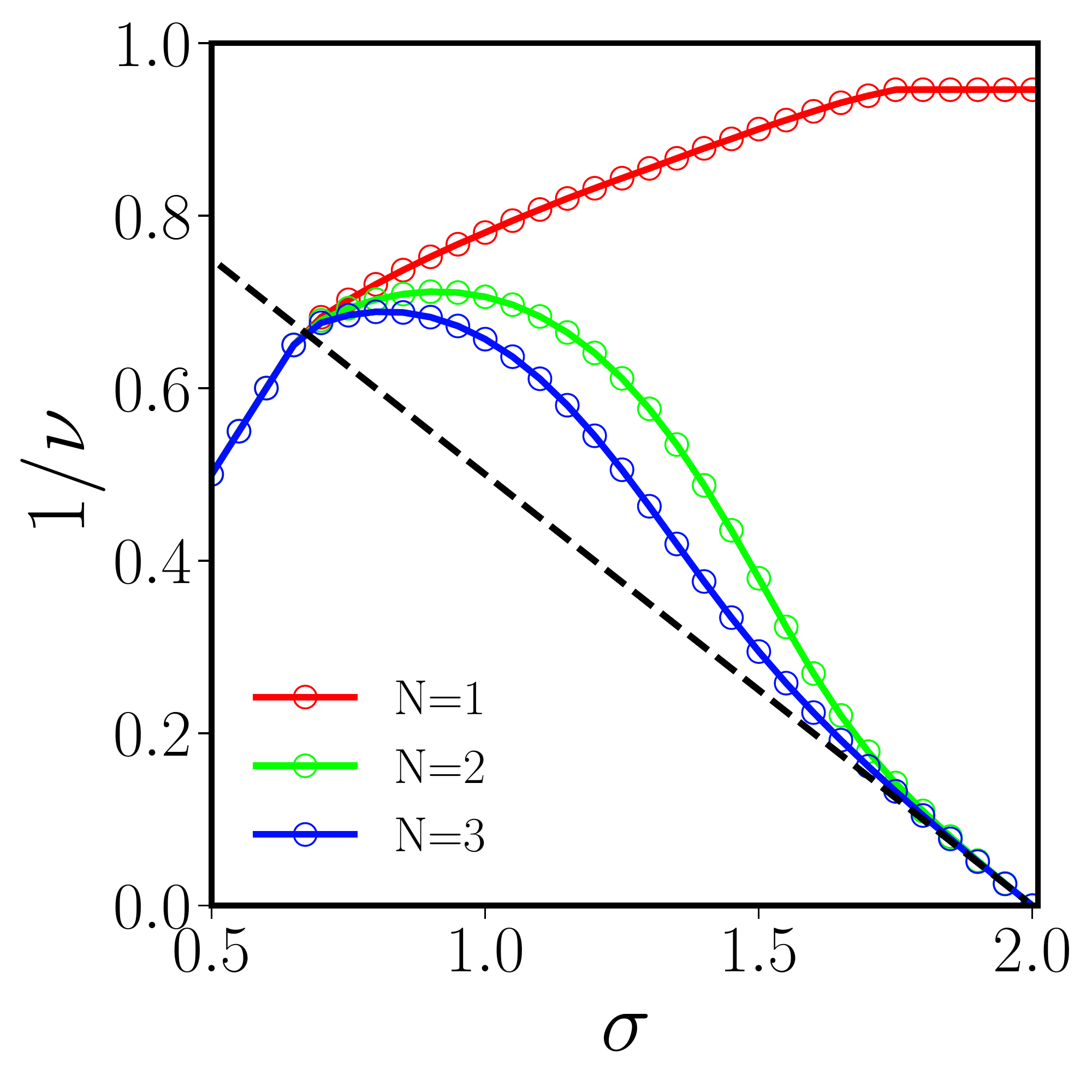}
     \caption{}
     \label{Fig5b}
    \end{subfigure}
\caption{\label{Fig5} The critical exponents $(z\nu)^{-1}$, panel (a), and $\nu^{-1}$, panel (b), as a function of $\sigma$ in $d=1$.
      The red, blue and green curves are for $N=1,2,3$ 
respectively, the mean field result is shown as a gray dashed line in panel (a). 
Again  in panel (a) 
the leading order term of a
$\tilde{\varepsilon}=(d-d_{lc})$ expansion is shown as a 
black dashed line. In panel (b) the black dashed line represents the $N\to\infty$ result.}
\end{figure}

Following Refs.\,\cite{Codello2015,Morris1998a}  the discrete spectrum of the eigenvalues $y_{t}$ is identified by the large field behaviour of the perturbing potentials $u(\bar{\rho})$. This spectrum only features a single positive eigenvalue, which describes the scaling of the relevant perturbation at the second order QCP, and it coincides with the inverse correlation length exponent ${\rm max}\,\, y_{t}=\nu^{-1}$.

In Fig.\,\ref{Fig5}  the numerical curves for $(z\nu)^{-1}$ and $\nu$ are reported in the $d=1$ case, panels\,\ref{Fig5a}  and\,\ref{Fig5b} respectively. 
The comparison with the MC results reported Ref.\,\cite{Sperstad2012} shows remarkable agreement for the $\mathbb{Z}_{2}$ symmetry case $N=1$. 

On the other hand, the $O(2)$ case (blue solid lines in Fig.\,\ref{Fig5}) is in disagreement with the numerical findings of Ref.\,\cite{Sperstad2012} for an XY spin chain with non-Ohmic dissipation. 
However, MC simulation are more difficult
close to the topological phase transition appearing in the $\sigma \to 2$ limit, due to the presence of logarithmic finite size corrections. Indeed, the BKT physics expected in this case predicts an exponential 
divergence of the correlation length ($\nu=\infty$), as it is proven by the generalisation of the  $\tilde{\varepsilon}=d-2$ expansion discussed Ref.\,\cite{Brezin1976} to the present case (dashed black line in Fig.\,\ref{Fig5}), see also Ref.\,\cite{defenu17} for an explanation on how to derive this expansion using the effective dimension approach. 

Finally, the overall picture for the universal behaviour of LR quantum $O(N)$ rotor models can be obtained considering the analytical results $\eta=2-\sigma$, together with the numerical results for $1/(z\nu)$ and $1/\nu$ shown in Fig.\,\ref{Fig5}. All the remaining critical exponents can be obtained by scaling relations. Explicit numerical values for the critical exponents have been given in Ref.\,\cite{defenu17}\footnote{Please note that a misprint is present in Ref.\,\cite{defenu17}, where in Tab.\,I it is written $d=2,3$ instead of $d=1,2$.}.

\section{Conclusions}
\label{Sec5}
In this work we defined LR interactions an interacting potential $V(r)\propto r^{-d-\sigma}$, which is  power-law decaying with the distance $r$ between two of the system microscopic components. The parameters  $d$ and $\sigma$ indicate respectively the dimension of the system and a positive power-law shift which makes the interaction energy finite. As a function of $\sigma$ the many body systems with LR interactions present a wide range of different physical behaviors. 
For $\sigma<0$ the interaction energy may blow up in the thermodynamic limit, and the traditional thermodynamic description may not be well defined. 
Then, it is necessary to rescale the
interaction strength with the system size\,\cite{Kac1963} in order to obtain a finite internal energy, but several properties of these systems will still violate the traditional thermodynamic picture\,\cite{Campa2014}. 

Our analysis was focused on the $\sigma>0$ 
where the system is additive and thermodynamics is well defined. 
Both in the classical and in the quantum case we have identified three regions in
the $(d,\sigma)$ phase diagram of the model. If the system dimension is larger than the upper critical dimension $d > d_{{\rm uc}}$, where $d_{{\rm uc}}$ is a $\sigma$ dependent value the system belongs to the mean-field universality class,
while for $\sigma$ above a threshold value $\sigma_{*}$ the system has the same critical behaviour of its SR analogue. Our studies then focused 
on the region $ \sigma \le \sigma_{*}$ and $d<d_{{\rm uc}}$ where LR interactions are relevant and the SSB transition belongs to a peculiar LR universality. 

In the first part of this paper we focused on classical  long-range (LR) $O(N)$ models 
in dimension $d\ge2$.
Using the flow equation for the effective potential
alone we found 
that universality classes of $O(N)$ LR models are 
in correspondence with those of $O(N)$ short range (SR)  
models in effective dimension $D_{\rm eff}=2d/\sigma$.
We also found new multicritical potentials which are present, at fixed $d$, 
above certain critical values of the parameter $\sigma$. 
 
We then investigated anomalous dimension effects by considering the flow of a field independent wave function renormalization. 
We found $\delta\eta=0$, i.e. the Sak's result\,\cite{Sak1973} 
in which there are no correction to the mean--field value 
of the anomalous dimension. 
The relation between the LR 
and the SR models remains valid at this approximation level using the effective dimension $D_{\rm eff}'$ 
defined by Eq.\,\,\eqref{EffectiveDimensionLPA$'$},
while the correlation length exponent is given according 
to Eq.\,\eqref{NuRelationLPA2}. 
Quantitative predictions for the exponent $\nu_{LR}$ for various values of 
$N$ were as well presented in $d=2$, see Fig.\ref{Fig1}.

Finally we introduced an effective action
where both the SR and LR terms
are present in the propagator. This approach does not impose {\it a priori} 
which is the dominant coupling in the RG flow. 
We showed how Sak's result is justified 
by the fixed point structure of the model, 
where a LR interacting fixed point appears only 
if $\sigma<\sigma_*$ and it controls the critical behaviour of the system, see the stability matrix in Fig.\,\ref{Fig2}. 
 An important result is that the effective dimension $D_{\rm eff}'$ can be shown 
to be not exact 
at this  improved approximation level and  it is 
possible to estimate the error present using the 
effective dimension $D_{\rm eff}'$. This error is found to be
rather small even close to the $\sigma_{*}$ threshold. Given the fact that often 
models with an effective dimension are used to take into account effects 
of long ranginess, competing interactions and disorder, we hope that our 
analysis based on the ansatz in  Eq.\,\eqref{EffectiveAction_LPA$''$} and the estimate 
of the error done for the critical exponents may call 
for similar analyses in a variety of other physical systems where 
the effective dimension is introduced. 

The final picture emerging from our analysis is the following: starting at 
$\sigma=0$ and increasing $\sigma$ towards $2$, we have found that for $\sigma<d/2$ 
only the LR Gaussian fixed  point exists and no SR terms in the fixed point propagator are 
present. At $\sigma=d/2$ a new interacting 
fixed point emerges from the LR Gaussian one and the same happens at the 
values  
$\sigma_{c,i}$ where new LR universality classes appear (in the same way as 
the multicritical SR fixed points are generated below the upper critical 
dimensions). Finally, when $\sigma$ approaches $\sigma_*$ the LR Wilson--Fisher 
fixed point merges with its SR equivalent and the LR term in the propagator 
disappears for $\sigma>\sigma_*$: this 
has to be contrasted with the case $\sigma<\sigma_*$ 
where at the interacting LR fixed points the propagator 
contains also a SR term. 
The same scenario is valid for all multicritical 
fixed points, provided that the $\sigma_*$ values  are computed 
with the corresponding SR anomalous dimensions.

The reliability of our approach is not spoiled by any change in the system external parameters $d$ and $N$ and it can also be employed to study multi-critical universalities. Therefore, the picture obtained in the present paper, see also Ref.\,\cite{Defenu2015}  encompasses the whole zoology of SSB transitions occurring in ferromagnetic $O(N)$ models and clarifies the controversy on the existence of non mean-field corrections to the LR anomalous dimension\,\cite{Sak1973,Picco2012,Blanchard2013,Brezin2014}. 
The agreement with MC simulations is remarkably good, especially considering the logarithmic corrections that plague the model for $\sigma\simeq\sigma_{*}$, as it has been confirmed on the Ising model\,\cite{Luijten2002,Angelini2014,Mori2010,Horita2016}, but also, 
using effective dimension relations,
on $1$-dimensional LR bond percolation\,\cite{Gori2016}. 

Having constructed a consistent picture for the physics of classical LR $O(N)$ models in the first part of the present paper, the second part was focused on the extension of the aforementioned picture to the quantum case. An important reason for the interest in the equilibrium and non-equilibrium 
properties of quantum LR systems is the connection with typical themes 
of classical LR physics. Therefore, one can ask: ``what are the phenomena introduced by LR interactions in classical systems which also survive in the quantum case?" and also ``What are the phenomena which are solely due to the interplay between quantum fluctuations and LR interactions?".
The study of LR interactions in quantum
systems has a long tradition in condensed matter physics. 
Indeed, the $\frac{1}{r^{2}}$
Ising model\,\cite{Thouless1969} has been shown to be related to the spin-$\frac{1}{2}$ Kondo problem, leading to   the occurrence of a 
topological phase transition of the Berezinskii-Kosterlitz-Thouless type\,\cite{Yuval1970,Cardy1981}.
Moreover, in the $90$'s the investigation of quantum LR 
Ising and rotor models pursued in Ref.\,\cite{Dutta2001} has been triggered by the realisation of dipolar Ising spin glasses.
In this framework our FRG study of these models extended previous investigations beyond the traditional perturbative framework and clarified how the appearance of finite anomalous dimension corrections to the critical propagator spoils the traditional quantum to classical correspondence in presence of LR interactions.

The effect of LR interactions on the system universal behaviour can be grasped by a simple scaling analysis of the critical propagator, which in the SR case obeys the asymptotic relation $G^{-1}(q)\propto q^{2}$ in the $q\to 0$ limit. When such relation is valid, the zero temperature QCP belong to the same universality class of the classical
phase transition, occurring at $T\neq 0$ in $d+1$ dimensions\,\cite{Sachdev2011}. 
As detailed in the text, LR interactions deeply alter the behaviour above, with the critical propagator scaling as $G^{-1}(q)\propto q^{\sigma}$. Therefore, 
the application of the Trotter decomposition to LR models leads to an
anisotropic field theory in the extended $d+1$ dimensional space. This anisotropy in the spatial and time coordinates spoils the traditional quantum to classical correspondence since the momentum and frequency kinetic sectors renormalise independently one form the other. Only for large enough $\sigma$ LR interactions become irrelevant and the isotropy in the $d+1$ space is restored.

The exactness of Trotter decomposition guarantees that the relation between the quantum model with its $d+1$ classical equivalent both isotropic and anisotropic, in the classical and quantum cases respectively, is exact. On the other hand, a further correspondence emerges from Eq.\,\eqref{Eq25}  between the quantum LR system in dimension $d$ and its classical analogous in dimension $d+z$, which can be also argued by simple scaling arguments, see Ref.\,\cite{Sachdev2011}. 
However, this $d\to d+z$ correspondence seems an artefact of our approximation procedure rather than an exact result.

The approximate nature of the $d\to d+z$ relation can be proved by the emergence of different coefficients in the evolution equation for the kinetic sectors in the anisotropic model in dimension $d+1$ and the isotropic one in dimension $d+z$.
However, the influence of these coefficients in practical computation is negligible and the the results depicted in Fig.\,\ref{Fig5} very well satisfy the equivalence between quantum $d$ dimensional system and 
classical $d+z$ one\,\cite{Defenu2015}.

The phase diagram of a LR system can be then 
represented in the $(d,\sigma)$ plane, as in Fig.\,\ref{Fig3}. For small 
enough values of the decay exponent $\sigma$ 
the system undergoes spontaneous symmetry breaking with mean field 
exponents
given by relations \eqref{eta_mf}, \eqref{z_mf} and \eqref{nu_mf}, green shaded region in Fig.\,\ref{Fig3}. For intermediate values
of $\sigma$ the system has peculiar LR exponents which continuously merge with the SR values at the $\sigma_{*}$ threshold.
In analogy with the classical case \cite{Angelini2014,Defenu2015},  
such threshold value
is given by $\sigma_{*}=2-\eta_{\rm{SR}}$, where $\eta_{\rm{SR}}$ 
is the anomalous dimension
found in the pure SR system.
It is worth noting that for continuous symmetries $N\geq 2$ it is 
also possible to identify a region
where no phase transition is possible. Such identification does not 
have a counterpart in the discrete symmetry case of 
the Ising $N=1$, where no exact lower critical dimension is 
known even in the traditional SR case. To the best of our knowledge the estimations for the critical exponents of quantum LR O(N) models presented in Figs.\,\ref{Fig4} and \,\ref{Fig5} are, at present, the most accurate in literature. 

Several experimental realisations of LR quantum systems are currently ongoing, but a precise quantitative determination of the universal properties for these systems is still missing. We hope that in the future our results may be used to benchmark and understand future experiments. The necessity to understand these phenomena is strongly tied with possible application of LR interacting systems to the construction of quantum information devices.

Finally, the occurrence of the 
BKT phase transition in the $d=1$, $N=2$ limit could not be properly addressed within the present formalism. However, our recent development in the construction of a reliable FRG formalism for the BKT transition\,\cite{DefenuBKT} make us confident on the feasibility of this study, especially in connection with the properties 
of the $1$-dimensional XXZ with LR couplings 
\cite{Laflorencie2005,Bermudez2016}. 

\textbf{Acknowledgements:} We acknowledge stimulating discussions with Alessandro Campa, Tilman Enss, Giacomo Gori, Shamik Gupta, Micheal Kastner and Giovanna Morigi. N.D. acknowledges financial support by Deutsche
Forschungsgemeinschaft (DFG) via Collaborative Research Centre SFB 1225 (ISOQUANT) and under Germany’s Excellence Strategy EXC-2181/1-390900948 (Heidelberg STRUCTURES Excellence Cluster).

\bibliography{proc}
\end{document}